\documentclass[aps,prb,tightenlines,twocolumn,superscriptaddress]{revtex4-1}
\usepackage{amsmath}
\usepackage{amsfonts}
\usepackage{amssymb}
\usepackage{color}

\usepackage{comment}

\usepackage{amsmath}
\usepackage{dcolumn}
\usepackage{multirow}
\usepackage{longtable}
\usepackage{bm}
\usepackage{slashed}
\usepackage[toc,page]{appendix}
\usepackage{float}
\usepackage{bbm}
\usepackage[pdftex]{graphicx}

\renewcommand{\dag}{^{\dagger}}
\newcommand{\bk}{\mathbf{k}}
\newcommand{\br}{\mathbf{r}}

\newcommand{\bG}{\mathbf{G}}
\newcommand{\bt}{\mathbf{t}}
\newcommand{\bR}{\mathbf{R}}
\newcommand{\bx}{\mathbf{x}}
\newcommand{\cA}{\mathcal{A}}
\newcommand{\cC}{\mathcal{C}}
\newcommand{\cF}{\mathcal{F}}
\newcommand{\cK}{\mathcal{K}}
\newcommand{\cP}{\mathcal{P}}
\newcommand{\cT}{\mathcal{T}}
\newcommand{\cW}{\mathcal{W}}
\newcommand{\bbZ}{\mathbbm{Z}}
\newcommand{\id}{\mathbbm{1}}
\newcommand{\ket}[1]{\left| #1 \right\rangle}
\newcommand{\bra}[1]{\left\langle #1 \right|}
\newcommand{\pars}[1]{\left( #1 \right)}
\newcommand{\brac}[1]{\left\{ #1 \right\}}
\newcommand{\eqn}[1]{\begin{equation} #1 \end{equation}}
\DeclareMathOperator{\dee}{d\!}
\DeclareMathOperator*{\Tr}{Tr}

\begin{document}
\title{Space group constraints on weak indices in topological insulators}
\author{D{\'a}niel Varjas}
\affiliation{Department of Physics, University of California, Berkeley, California 94720, USA}
\author{Fernando de Juan}
\affiliation{Department of Physics, University of California, Berkeley, California 94720, USA}
\affiliation{Instituto Madrile\~no de Estudios Avanzados en Nanociencia (IMDEA-Nanociencia), 28049 Madrid, Spain}
\author{Yuan-Ming Lu}
\affiliation{Department of Physics, The Ohio State University, Columbus, Ohio 43210, USA}
\bibliographystyle{apsrev4-1}
\begin{abstract}
Lattice translation symmetry gives rise to a large class of ``weak'' topological insulators (TIs), characterized by translation-protected gapless surface states and dislocation bound states. In this work we show that space group symmetries lead to constraints on the weak topological indices that define these phases. In particular we show that screw rotation symmetry enforces the Hall conductivity along the screw axis to be quantized in multiples of the screw rank, which generally applies to interacting systems. We further show that certain 3D weak indices associated with quantum spin Hall effects (class AII) are forbidden by the Bravais-lattice and by glide or even-fold screw symmetries. These results put a strong constraints on candidates of weak TIs in the experimental and numerical search for topological materials, based on the crystal structure alone.



\end{abstract}
\maketitle

\section{Introduction}




The discovery of topological insulators and superconductors is one of the most important breakthroughs of condensed matter physics in the past decades\cite{Hasan2010,Hasan2011,Qi2011}. The key principle underlying the existence of these novel topological phases is that the presence of a symmetry, such as time reversal symmetry ($\cT$), can lead to a quantized bulk topological invariant and robust gapless surface states. In a gapped fermion system, this invariant cannot change unless the gap closes, defining a stable quantum phase and protecting the existence of gapless boundary states. After the discovery of three dimensional topological insulators, which are protected by ($\cT$), it was realized that other global symmetries in the Altland-Zirnbauer (AZ) classes\cite{AZ}, such as charge conjugation ($\cC$) and spin rotational symmetries, also give rise to topological phases, leading to the periodic table\cite{Schnyder2008,Kitaev} of topological insulators and superconductors.

It was realized early on that additional topological phases can be obtained from invariants defined on a lower dimensional slice of the Brillouin Zone \cite{BalentsMoore} (BZ). Since this definition requires the discrete translational symmetry of the lattice, it was initially thought that these phases would not survive generic disorder and thus termed ``weak'' topological insulators. The lower-dimensional topological invariants are therefore known as weak indices. However, further efforts then showed that weak topological phases have many robust features emerging from topology\cite{MBM12,RKS12,YMT13,MF14}, in particular the fact that lattice dislocations host protected gapless modes that originate from the weak indices\cite{Ran,Ran10}. Recently it was also proposed that strong interactions can lead to novel topological orders on the surface of weak TIs\cite{Qi2015,Mross2016}. Most of these theoretical predictions remain untested due to the difficulty of finding materials realizing these weak topological phases\cite{RIR13}, though several candidates have been predicted in ab-initio studies\cite{YMF12,TYC14,YLF14,LZY16}

The consideration of a perfect lattice with translational symmetry immediately raises the question of whether the space group symmetries of this lattice may also have an impact on the topological properties. The addition of space group often leads to the emergence of novel phases, generally termed topological crystalline insulators\cite{FuKaneInversion,BernevigPGS,HughesPGS,Zaanen,Chiu,Morimoto,Lu,FuGlide,Shiozaki,Mong,LiuNS}, with different properties from weak TIs. Here we address a complementary question: what are the restrictions brought by space group symmetries on possible topological phases, in particular, the weak topological phases?  

In this work, we show that the non-symmorphic elements of the space group lead to unexpected constraints for the weak indices beyond those derived from the point group. First, we show that for three dimensional (3D) magnetic insulators in class A there is nontrivial quantization condition of Hall conductivity tensor in the presence of non-symmorphic screw symmetry. We derive this condition from band theory, and then provide a general proof of its applicability to interacting systems. Second, we turn to time-reversal-invariant insulators in class AII and show how nonsymmorphic screw and glide symmetries can make the weak indices vanish in a particular direction. While enumerating every AZ symmetry class and dimensionality is beyond the scope of this paper, we present the necessary formalism to generalize our results to topological superconductors with a few examples in Appendix~\ref{sec:TSC}.

\section{Chern number and Hall conductivity (class A)}

\subsection{Hall conductivity of a 3D insulator}

A simplest example of weak indices in a three dimensional system is the quantized Hall conductivity of an insulator, which in proper units is given by integer-valued Chern numbers of 2D slices of the BZ. As off-diagonal elements of the conductivity tensor, these Chern numbers transform like an axial vector under point group operations. Here we show that a non-symmorphic screw symmetry further imposes an important constraint on the integer-valued Hall conductivity. This constraint holds generally for interacting and disordered systems, as long as the ground state is a non-fractionalized 3D insulator which preserves the screw symmetry.

In a 2D system, the Hall conductance (or conductivity) $\sigma_{xy}$ characterizes the transverse current response to an in-plane electric field: $j_x = \sigma_{xy} E_y$. Using the Kubo formula one finds\cite{Thouless1982,Nagaosa2010} that the Hall conductivity in unit of $e^2/h$ is given by the integral of Berry curvature for the occupied bands over the BZ:
\eqn{
\sigma_{xy} =\frac{e^2}{h}\mathcal{C},~~~\mathcal{C}=\frac{1}{2\pi}\int_{\textnormal{BZ}}\text{d}^2{\bf k} \Tr_{occ.} \cF_{\bf k}
}
where $\Tr_{occ.}$ is the trace over occupied bands and $\cF$ is the Berry curvature matrix. In an insulator with a bulk gap between valence and conduction bands, the total Berry flux over the BZ is quantized to be an integer, known as the Chern number $\mathcal{C}$, and hence $\sigma_{xy}$ is also quantized.

In a 3D insulator the Hall conductivity becomes an antisymmetric tensor and can be cast in terms of an axial vector\cite{Nagaosa2010} $\mathbf{\Sigma}$ in the form of
\eqn{\label{eqn:hall conductivity tensor}
\sigma_{ij} = \frac{e^2}{2\pi h}\epsilon_{ijl}\Sigma_l,~~~\Sigma_i=\frac{\epsilon_{ijl}}{4\pi}\int_{\textnormal{BZ}}\text{d}^3{\bf k} \Tr_{occ.} \cF^{jl}_{\bf k}
}
where repeated indices are summed over implicitly. In band insulators this ``Hall vector'' is always a reciprocal lattice vector\cite{Halperin1987} and can be expressed as $\mathbf{\Sigma} =\sum_{i=1}^3\bG_i\mathcal{C}_i$ where $\bG_i$ are an independent set of primitive reciprocal lattice vectors, and $\mathcal{C}_i\in\bbZ$ is the Chern-number for a cut of the BZ oriented towards $\bG_i$ and spanned by the other two reciprocal lattice vectors. The weak topological invariant associated with 3D insulators in symmetry class A is such a ``Chern vector'' $\vec{\mathcal{C}}\in\bbZ^3$. In Appendix~\ref{sec:trfproof} we prove that the ``Hall vector'' vector transforms as an axial vector even for nonsymmorphic symmetries.  This shows that lattice symmetry severely constrains the allowed values, as it has to stay invariant under every orthogonal transformation in the point group. Nonzero values are only allowed with low enough symmetry, two intersecting axes of (improper) rotations is sufficient to force vanishing Hall conductance\citep{BernevigPGS}.

For a finite 3D insulator with lattice translational symmetry, the Hall conductance normal to a certain direction $\hat m$ is given by
\eqn{
\epsilon_{mnl}\sigma_{nl}=\frac{e^2}{2\pi h}\mathbf{\Sigma}\cdot(L_m\hat m)=\frac{e^2}{h}\sum_{i}\frac{L_m\hat m\cdot\bG_i}{2\pi}\mathcal{C}_i
}
where $L_m$ is the linear size of the periodic system along $\hat m$ direction. If this 3D system has a gapped bulk, its Hall conductance must also be an integer in units of $e^2/h$ along any direction. Adding one extra unit cell along $\bG_i$ direction will increase the Hall conductance along $\bG_i$ direction by exactly the Chern number $\mathcal{C}_i$ in units of $e^2/h$. In an anisotropic limit the 3D insulator can be viewed as a stack of 2D layers  with a quantized Hall conductance $\sigma_{xy}^L$ each.  Therefore the Hall conductivity tensor (\ref{eqn:hall conductivity tensor}) is nothing but the Hall conductance per unit cell layer, $\sigma_{xy}^L$, which can be defined as the difference between the Hall conductance of $N_z$ and $N_z + 1$ layers along a certain direction. As will become clear later, this difference $\sigma_{xy}^L$ does not depend on $N_z$ as long as $N_z$ is much larger than the correlation length, so we adopt this definition for our interacting proof.

In the following we show that, with a non-symmorphic $n$-fold screw symmetry, the Hall conductance per unit cell layer along the screw-axis direction cannot be an arbitrary integer (in units of $e^2/h$) for a gapped 3D insulator without fractionalization. Instead it must be a multiple of $n$, as enforced by the screw symmetry.

%

\subsection{Screw symmetry enforced constraints}
\begin{figure}[tb]
\begin{center}
\includegraphics[width=8.5cm]{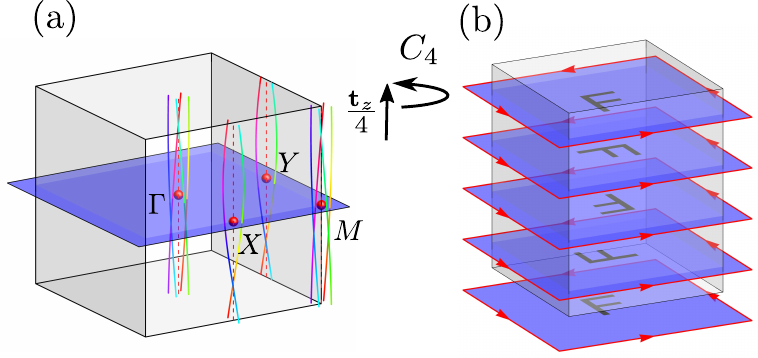}
\caption{(a) Brillouin zone of an insulator with 4-fold screw symmetry. The perpendicular plane through the BZ center (blue) contains four high symmetry points (red spheres), we use the formula relating the chern number to rotation eigenvalues at these 4 points. In the direction of the screw, along the four invariant lines (dashed red) the screw eigenvalues evolve into each other, illustrated by spirals and color code. Note that at lines that are only invariant under twofold screw there are only two different eigenvalues, but the total number of occupied bands is still a multiple of 4. (b) Intuitive real space picture of the screw symmetric insulator as a stack of integer Chern insulator layers related by the screw. The unit cell contains four layers, so the Hall conductance per transverse unit cell is a multiple of 4.}\label{fig:screwBZ}
\end{center}
\end{figure}
Consider an essential $n$-fold screw in the $z$ direction. Below we will show the Chern number for a cut perpendicular to the screw direction is quantized to a multiple of $n$. By essential screw we mean a SG operation that leaves no point in space invariant up to lattice translations\cite{SidNS}. In an appropriately chosen basis the translation part is $1/n$ of the primitive lattice vector parallel to the $n$-fold rotational axis, $g=\brac{C_n \middle| \mathbf{a}_z /n}$. For this we invoke results\cite{BernevigPGS} (also see Appendix~\ref{sec:Cn}) that allow calculation of the Chern number in the presence of $n$-fold rotational symmetry in 2D as a product of rotation eigenvalues of occupied bands at high symmetry points of the BZ. For example with $C_4$ symmetry (Fig.~\ref{fig:screwBZ}~(a))
\eqn{
\exp\pars{2\pi i\frac{C}{4}} = \prod_{m\in\textnormal{occ.}} \xi_m^{\Gamma}\pars{C_4} \xi_m^{M}\pars{C_4} \xi_m^{X}\pars{C_4^{-2}}
\label{eqn:C4}
}
where $\xi_m^{\bk}\pars{O}$ is the rotation eigenvalue of $O$ in band $m$ at momentum $\bk$. Similar formulae can be derived for rotations $C_2$, $C_3$ and $C_6$. 

When restricted to the 2D cut of the BZ through $\Gamma$, a screw acts the same way as a symmorphic $C_n$ rotation, so the formula can be applied. Now consider the screw-invariant lines in BZ, parallel to the screw axis (dotted vertical lines in FIG. \ref{fig:screwBZ}). As the $n$-th power of the screw $g^n = (-1)^F \brac{\id\middle| \mathbf{a}_z}$ is a pure translation up to fermion parity, the eigenvalues of screw $g$ take values of $\exp\left(i \bk\cdot\mathbf{a}_z/n + 2\pi i m/n + \pi i F/n\right)$ for $m \in \mathbbm{Z}_n$. When restricted to the perpendicular plane with $\bk\cdot\mathbf{a}_z = 0$ the eigenvalues are simply the $n$-th roots of fermion parity $(-1)^F$. Increasing $k_z$ by $2\pi$ will change $g$ eigenvalue by a factor of $e^{2\pi i/n}$, leading to a $n$-multiplet of occupied bands at each screw-invariant momentum (Fig.~\ref{fig:screwBZ}~(a)). 
%
This shows that the product of screw eigenvalues at high-symmetry points is always 1. This immediately proves that 
\eqn{\label{eqn:n-fold screw:constraint}
\left. \frac{\sigma_{xy}^L}{{e^2}/{h}}\right. = \mathcal{C}_z \equiv 0 \mod{n}.
}
\begin{figure}[tb]
\begin{center}
\includegraphics[width=8.5cm]{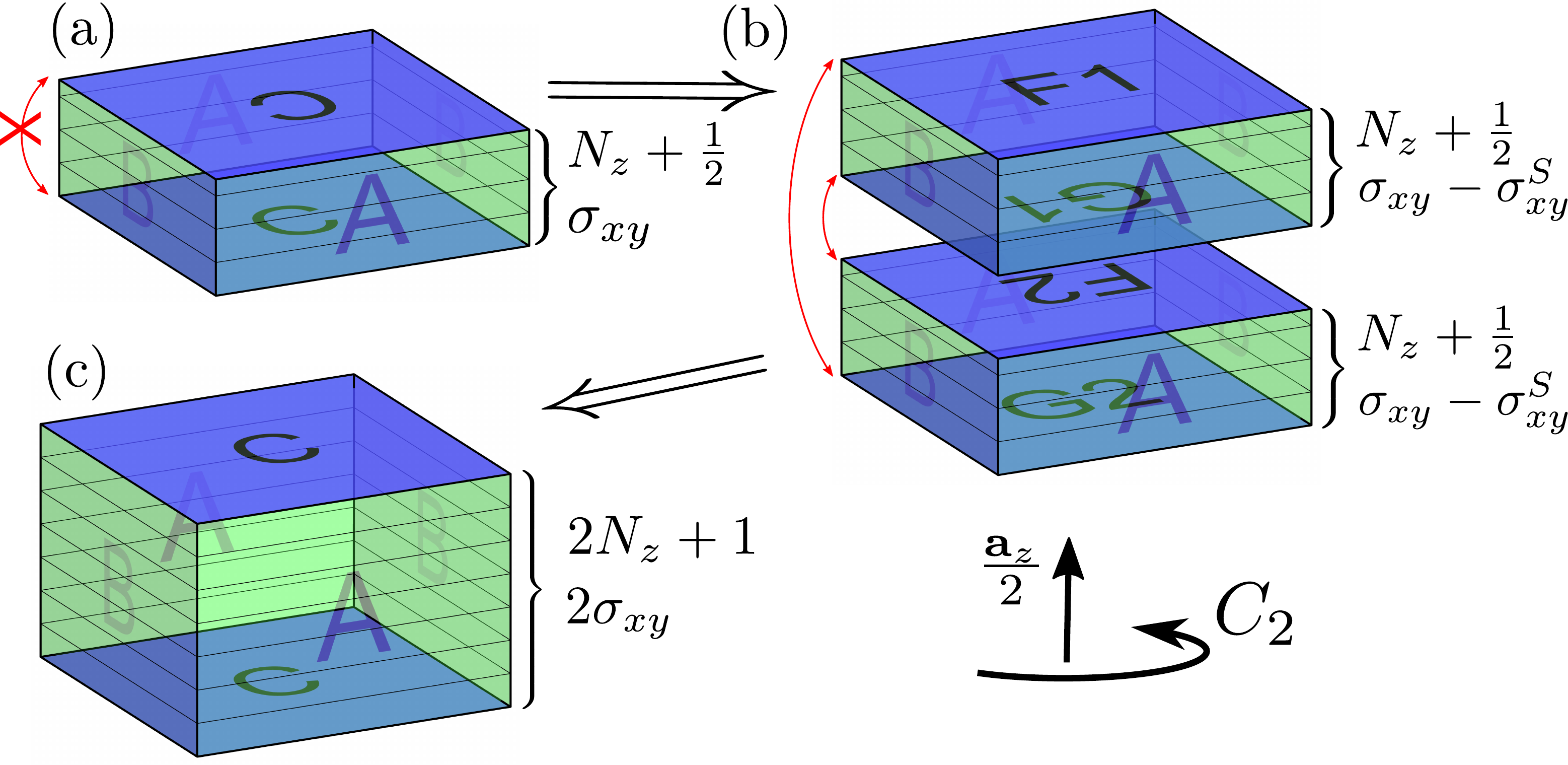}
\caption{The process used in the general proof for the Hall conductance constraint (\ref{eqn:n-fold screw:constraint}), illustrated in the case of a 2-fold screw. (a) A thick slab with half-integer thickness and twisted periodic boundary condition in the $z$ direction. As we open the boundary condition in the $z$ direction, the Hall conductance may change by a surface contribution $-\sigma_{xy}^S$. (b) We combine two slabs, the screw axis allows to arrange these such that the interfaces are guaranteed to be identical, a top surface (F) meets a bottom surface (G) with the same orientation. Glueing the two interfaces together each contributes $+\sigma_{xy}^S$. (c) The resulting system has periodic boundary conditions in all three directions with odd thickness while the Hall conductance is an even multiple of the conductance quantum.}\label{fig:screwint}
\end{center}
\end{figure}
In the following we show that this result is not a peculiarity of band theory for free electrons, but holds for any gapped unique ground state preserving $n$-fold screw symmetry, even in the presence of interactions and disorders. The proof is based on the following cut and glue procedure. We start with a slab containing $N_z + m/n$ unit cells along the $z$ direction which is parallel to the screw axis. While this number of unit cells is not integer, screw symmetry allows us to identify the top and bottom surfaces using a boundary condition twisted by a $C_n$ rotation\cite{WatanabeNS}, which results in a screw symmetric bulk without boundaries (Fig.~\ref{fig:screwint}~(a)). To take the thermodynamic limit we assume the size of the system is much larger than the correlation length of the gapped bulk. To define the Hall conductance $\sigma_{xy}$ in this geometry we invoke the Streda formula\cite{Streda1982}, whereby the Hall conductance is given by the charge bound to a localized $2\pi$ flux threaded through the system. Unless the charge captured is an integer, the system is fractionalized and has a non-unique ground state, contradicting our initial assumption.

Next we cut the system open in the $z$ direction. During this process we change the Hall conductance by a surface contribution of $-\sigma_{xy}^S$. $\sigma_{xy}^S$ can depend on the thickness, but should saturate to a thickness-independent constant, as long as $N_z$ is much larger than the correlation length of the gapped bulk. We then take $n$ copies of this open system and arrange them along the $z$ direction related by $C_n^m$ rotations so that all the interfaces are symmetry related (Fig.~\ref{fig:screwint}~(b)). Gluing the surfaces together by restoring the screw symmetric bulk Hamiltonian changes the Hall conductance by $\sigma_{xy}^S$ at each interface, as the separation between them is much larger than the bulk correlation length. The resulting system (Fig.~\ref{fig:screwint}~(c)) has periodic boundary conditions in all three directions with a thickness of $nN+m$ unit cells and Hall conductance of $n\pars{\sigma_{xy}-\sigma_{xy}^S} + n \sigma_{xy}^S = n\sigma_{xy}$. Thus we proved that a sample with arbitrary integer thickness has a Hall conductance which is a multiple of $n$ times the conductance quantum.

\section{Weak TI indices (class AII)}
\subsection{Bravais lattice}
Point group symmetry can put stringent constraints on the allowed values of the weak indices. To calculate the weak $\bbZ_2$ invriants we evaluate\cite{BalentsMoore}
\eqn{
\nu_i = \frac{1}{2\pi} \Tr_{occ.}\pars{\int_{\frac{1}{2} T^2} \cF_{\bf k}\text{d}^2{\bf k} - \oint_{\partial \frac{1}{2} T^2} \cA_{\bf k}\cdot\text{d}{\bf k} }\mod{2}
\label{eqn:Z2}
}
where $\cA$ and $\cF$ are the Berry curvature and connection and the integral is over the interior and boundary of half of the time-reversal invariant 2D cut of the BZ spanned by the two reciprocal lattice vectors other than $\bG_i$ and displaced from the $\Gamma$ point by $\bG_i/2$. This defines a $\bk$-space vector  characterizing the weak indices, pointing to one of the 8 TR invariant momenta:
\eqn{
\bG^{\nu} = \frac{1}{2} \nu_i \bG_i.
}
This vector transforms under space group operations as $\bk$-space vectors (see Appendix \ref{sec:trfproof}). One can enumerate the allowed values of $\bG^{\nu}$ by inspecting tables for Wyckoff-positions of the reciprocal space groups: $\bG^{\nu}$ can only take values at points with half-integer Miller indices that are invariant under the point group up to reciprocal lattice vectors. This means that a face-centered cubic lattice (for example the common cubic diamond structure\cite{FuKane2007}) can not support nontrivial weak indices without breaking point group symmetries. To see this, consider the BCC reciprocal lattice. It is easy to see that there is no maximal symmetry $\bk$-point other that the $\Gamma$ point that represents trivial weak indices. The constraints only depend on the type of Bravais-lattice\footnote{A priori the constraints depend on the \emph{arithmetic crystal class} that is determined by the point group (geometric crystal class) and its action on the translations (Bravais lattice). There are 73 arythmetic crystal classes in 3d in one-to-one correspondence to symmorphic space groups. The reciprocal space group has the same point group as the space group, but the reciprocal Bravais lattice may have different centering. Direct enumeration\cite{ITA,Aroyo2006,Aroyo2014} reveals that the result in fact only depends on the Bravais-lattice, this can be rationalized noting that a Bravais-lattice contains arithmetic crystal classes that have a large enough point group to force certain relations among lattice vectors by mapping them to each other. As $\bG^{\nu}$ takes values on the same reciprocal Bravais-lattice with half the spacing, the constraints are the same for every arithmetic crystal class in a Bravais-lattice.}, listed in Table~\ref{tab:weakindexshort}.

These results can also be rationalized from the band inversion point of view. To get a weak TI, we need an odd number of band inversions among the four TR invariant momenta located on one plane offset from the origin. But point group symmetry relates some of these TR invariant momenta, and band inversion can only occur simultaneously on all symmetry related points. It is easy to see for example that in the BCC reciprocal crystal TR invariant momenta are symmetry equivalent in such a fashion that there is an even number of related points in any of the offset planes, explaining the lack of nontrivial weak TI's. On the other hand, a strong TI is possible with any SG, because since the $\Gamma$ point is always of maximal symmetry, it is possible to have a band inversion only at the $\Gamma$ point (see Appendix~\ref{sec:strongTI}).
\begin{table}[htbp]
  \centering
  \caption{Constraints on possible weak indices based on space group symmetry in time reversal invariant insulators. The allowed values of $\bG^{\nu}$ are labeled as in ITA\cite{ITA}. ``all" means all the 8 possible values are allowed, while if only $\Gamma$ appears no nontrivial weak indices are allowed by symmetry.}
    \begin{tabular}{ccl}
    \toprule
    Crystal system & Centering & Allowed values of $\bG^{\nu}$ \\ \hline

    \multirow{1}{*}{Triclinic} & P & all \\ \hline
    \multirow{2}{*}{Monoclinic} & P     &  all \\
          & C & $\Gamma$, $Y$, $A$, $M$ \\ \hline
    \multirow{4}{*}{Orthorhombic} & P & all \\
          & C  & $\Gamma$, $Y$, $T$, $Z$ \\
          & I  & $\Gamma$, $X$ \\
          & F  & $\Gamma$, $Y$, $T$, $Z$ \\ \hline
    \multirow{2}{*}{Tetragonal} & P & $\Gamma$, $Z$, $M$, $A$ \\
          & I & $\Gamma$, $M$ \\ \hline
    \multirow{2}{*}{Trigonal} & P & $\Gamma$, $A$ \\
          & I & $\Gamma$, $T$ \\ \hline
    \multirow{1}{*}{Hexagonal} & P & $\Gamma$, $A$ \\ \hline
    \multirow{3}{*}{Cubic} & P & $\Gamma$, $R$ \\
          & I & $\Gamma$, $H$ \\
          & F & $\Gamma$ \\
 \botrule
    \end{tabular}%
  \label{tab:weakindexshort}%
\end{table}%
\subsection{Nonsymorphic symmetries}
The presence of non-symmorphic symmetries leads to further constraints on the weak indices. We now show that in the presence of an essential 2-fold screw in the $z$ direction, the weak index must be trivial in this direction. First we note that a 2-fold screw $\brac{C_2\middle|\mathbf{a}_z/2}$ squares to $\brac{-\id|\mathbf{a}_z}$, represented at $k_z = \pi$ as $+\id$, which commutes with $\cT$. Thus, in the $k_z = \pi$ plane it acts like a proper inversion in a 2D system and we can use the known result to evaluate the weak index by counting inversion eigenvalues\cite{FuKaneInversion}. The index is given by the product of the inversion eigenvalues at the four invariant momenta for each Kramers pair among occupied bands, note that in this plane Kramers partners have the same screw eigenvalues, as required.

However, this situation at $k_z=\pi$ in a 3D system is different from a 2D system with a symmorphic inversion symmetry in that the screw requires an equal number of both screw eigenvalues below the gap, as shown earlier. Specifically at $k_z = \pi$ at each high symmetry point the number of occupied  $+1$ and $-1$ eigenvalues must be equal, and as the total number of occupied bands is constant, the number of occupied $+1$ ($-1$) bands is a multiple of 4, leading to a trivial $\bbZ_2$ index in this plane.



To cast the statement in a more general form applicable to all lattices, consider that the above calculation requires a choice of basis in $\bk$-space where two of the primitive reciprocal lattice vectors are perpendicular to the screw axis. Such a choice is always possible, as for an essential 2-fold screw in a basis that removes any perpendicular translation, the translational part $\bt$ is half of a lattice vector, so there exists a set of primitive real space lattice vectors such that one of them is parallel to the screw axis, $\mathbf{a}_3 = 2\bt$. The corresponding reciprocal lattice vectors satisfy the above properties, and the statement is about the third component $\bG_3$ that has nonzero projection onto the screw axis. We need $\pars{e^{i \frac{1}{2} \bG_3\cdot \bt}}^2 = -1$ to get an effective inversion, in this case $\nu_3 = 1$ is forbidden. So a vector weak index is forbidden if
\eqn{
2 \pars{\bG^{\nu}\cdot \bt} = \pi \mod{2\pi}.
}

We analogously argue that an essential glide forbids nontrivial weak index in the direction parallel to the the translational part of the glide. Consider a glide translating in the $z$ direction and reflecting $x$, $\brac{m_x\middle|\mathbf{a}_z/2}^2 = \brac{-\id|\mathbf{a}_z}$, we find that at $k_z = \pi$ it acts like a mirror that squares to $+\id$ and commutes with $\cT$. This 2D system has trivial topological classification\cite{Morimoto,Chiu}, thus the weak index in this direction is zero. The general statement is that an essential glide that only contains in-plane translation $\bt$ forbids a weak vector if $2 \bG^{\nu}\cdot \bt = \pi \pmod{2\pi}$.

\section{Conclusion}
In summary, we derived a set of constraints on weak topological indices in 3D insulators from non-symmorphic and symmorphic point-group symmetries. We show that in the presence of $n$-fold screw rotation, Hall conductivity must be quantized as a multiple of $n$ for any 3D non-fractionalized insulator preserving screw symmetry. This condition is generally proved for interacting systems. We also show certain 3D weak indices for TIs (class AII) are forbidden by glide or even-fold screw symmetries. These results put strong constraints on the candidates for weak topological phases in the ongoing experimental and numerical efforts to find physical realizations of this novel topological phases.

The authors are grateful to Takahiro Morimoto and Hoi Chun Po for helpful conversations. This work is supported by NSF Grant No.~DMR-1206515 (DV and FdJ), the European Research Council Advanced Grant (Contract No. 290846) (FdJ) and startup funds at Ohio State University (YML).

\bibliography{nonsymmorphic}

\begin{thebibliography}{45}%
\makeatletter
\providecommand \@ifxundefined [1]{%
 \@ifx{#1\undefined}
}%
\providecommand \@ifnum [1]{%
 \ifnum #1\expandafter \@firstoftwo
 \else \expandafter \@secondoftwo
 \fi
}%
\providecommand \@ifx [1]{%
 \ifx #1\expandafter \@firstoftwo
 \else \expandafter \@secondoftwo
 \fi
}%
\providecommand \natexlab [1]{#1}%
\providecommand \enquote  [1]{``#1''}%
\providecommand \bibnamefont  [1]{#1}%
\providecommand \bibfnamefont [1]{#1}%
\providecommand \citenamefont [1]{#1}%
\providecommand \href@noop [0]{\@secondoftwo}%
\providecommand \href [0]{\begingroup \@sanitize@url \@href}%
\providecommand \@href[1]{\@@startlink{#1}\@@href}%
\providecommand \@@href[1]{\endgroup#1\@@endlink}%
\providecommand \@sanitize@url [0]{\catcode `\\12\catcode `\$12\catcode
  `\&12\catcode `\#12\catcode `\^12\catcode `\_12\catcode `\%12\relax}%
\providecommand \@@startlink[1]{}%
\providecommand \@@endlink[0]{}%
\providecommand \url  [0]{\begingroup\@sanitize@url \@url }%
\providecommand \@url [1]{\endgroup\@href {#1}{\urlprefix }}%
\providecommand \urlprefix  [0]{URL }%
\providecommand \Eprint [0]{\href }%
\providecommand \doibase [0]{http://dx.doi.org/}%
\providecommand \selectlanguage [0]{\@gobble}%
\providecommand \bibinfo  [0]{\@secondoftwo}%
\providecommand \bibfield  [0]{\@secondoftwo}%
\providecommand \translation [1]{[#1]}%
\providecommand \BibitemOpen [0]{}%
\providecommand \bibitemStop [0]{}%
\providecommand \bibitemNoStop [0]{.\EOS\space}%
\providecommand \EOS [0]{\spacefactor3000\relax}%
\providecommand \BibitemShut  [1]{\csname bibitem#1\endcsname}%
\let\auto@bib@innerbib\@empty
\bibitem [{\citenamefont {Hasan}\ and\ \citenamefont {Kane}(2010)}]{Hasan2010}%
  \BibitemOpen
  \bibfield  {author} {\bibinfo {author} {\bibfnamefont {M.~Z.}\ \bibnamefont
  {Hasan}}\ and\ \bibinfo {author} {\bibfnamefont {C.~L.}\ \bibnamefont
  {Kane}},\ }\href {http://link.aps.org/doi/10.1103/RevModPhys.82.3045}
  {\bibfield  {journal} {\bibinfo  {journal} {Rev. Mod. Phys.}\ }\textbf
  {\bibinfo {volume} {82}},\ \bibinfo {pages} {3045} (\bibinfo {year}
  {2010})}\BibitemShut {NoStop}%
\bibitem [{\citenamefont {Hasan}\ and\ \citenamefont
  {Moore}(2011)}]{Hasan2011}%
  \BibitemOpen
  \bibfield  {author} {\bibinfo {author} {\bibfnamefont {M.~Z.}\ \bibnamefont
  {Hasan}}\ and\ \bibinfo {author} {\bibfnamefont {J.~E.}\ \bibnamefont
  {Moore}},\ }\bibfield  {booktitle} {\emph {\bibinfo {booktitle} {Annual
  Review of Condensed Matter Physics}},\ }\href
  {http://dx.doi.org/10.1146/annurev-conmatphys-062910-140432} {\bibfield
  {journal} {\bibinfo  {journal} {Annu. Rev. Condens. Matter Phys.}\ }\textbf
  {\bibinfo {volume} {2}},\ \bibinfo {pages} {55} (\bibinfo {year}
  {2011})}\BibitemShut {NoStop}%
\bibitem [{\citenamefont {Qi}\ and\ \citenamefont {Zhang}(2011)}]{Qi2011}%
  \BibitemOpen
  \bibfield  {author} {\bibinfo {author} {\bibfnamefont {X.-L.}\ \bibnamefont
  {Qi}}\ and\ \bibinfo {author} {\bibfnamefont {S.-C.}\ \bibnamefont {Zhang}},\
  }\href {http://link.aps.org/doi/10.1103/RevModPhys.83.1057} {\bibfield
  {journal} {\bibinfo  {journal} {Rev. Mod. Phys.}\ }\textbf {\bibinfo {volume}
  {83}},\ \bibinfo {pages} {1057} (\bibinfo {year} {2011})}\BibitemShut
  {NoStop}%
\bibitem [{\citenamefont {{Altland}}\ and\ \citenamefont
  {{Zirnbauer}}(1997)}]{AZ}%
  \BibitemOpen
  \bibfield  {author} {\bibinfo {author} {\bibfnamefont {A.}~\bibnamefont
  {{Altland}}}\ and\ \bibinfo {author} {\bibfnamefont {M.~R.}\ \bibnamefont
  {{Zirnbauer}}},\ }\href {\doibase 10.1103/PhysRevB.55.1142} {\bibfield
  {journal} {\bibinfo  {journal} {\prb}\ }\textbf {\bibinfo {volume} {55}},\
  \bibinfo {pages} {1142} (\bibinfo {year} {1997})},\ \Eprint
  {http://arxiv.org/abs/cond-mat/9602137} {cond-mat/9602137} \BibitemShut
  {NoStop}%
\bibitem [{\citenamefont {Schnyder}\ \emph {et~al.}(2008)\citenamefont
  {Schnyder}, \citenamefont {Ryu}, \citenamefont {Furusaki},\ and\
  \citenamefont {Ludwig}}]{Schnyder2008}%
  \BibitemOpen
  \bibfield  {author} {\bibinfo {author} {\bibfnamefont {A.~P.}\ \bibnamefont
  {Schnyder}}, \bibinfo {author} {\bibfnamefont {S.}~\bibnamefont {Ryu}},
  \bibinfo {author} {\bibfnamefont {A.}~\bibnamefont {Furusaki}}, \ and\
  \bibinfo {author} {\bibfnamefont {A.~W.~W.}\ \bibnamefont {Ludwig}},\ }\href
  {\doibase 10.1103/PhysRevB.78.195125} {\bibfield  {journal} {\bibinfo
  {journal} {Phys. Rev. B}\ }\textbf {\bibinfo {volume} {78}},\ \bibinfo
  {pages} {195125} (\bibinfo {year} {2008})}\BibitemShut {NoStop}%
\bibitem [{\citenamefont {{Kitaev}}(2009)}]{Kitaev}%
  \BibitemOpen
  \bibfield  {author} {\bibinfo {author} {\bibfnamefont {A.}~\bibnamefont
  {{Kitaev}}},\ }in\ \href {\doibase 10.1063/1.3149495} {\emph {\bibinfo
  {booktitle} {American Institute of Physics Conference Series}}},\ \bibinfo
  {series} {American Institute of Physics Conference Series}, Vol.\ \bibinfo
  {volume} {1134},\ \bibinfo {editor} {edited by\ \bibinfo {editor}
  {\bibfnamefont {V.}~\bibnamefont {{Lebedev}}}\ and\ \bibinfo {editor}
  {\bibfnamefont {M.}~\bibnamefont {{Feigel'Man}}}}\ (\bibinfo {year} {2009})\
  pp.\ \bibinfo {pages} {22--30},\ \Eprint {http://arxiv.org/abs/0901.2686}
  {arXiv:0901.2686 [cond-mat.mes-hall]} \BibitemShut {NoStop}%
\bibitem [{\citenamefont {{Moore}}\ and\ \citenamefont
  {{Balents}}(2007)}]{BalentsMoore}%
  \BibitemOpen
  \bibfield  {author} {\bibinfo {author} {\bibfnamefont {J.~E.}\ \bibnamefont
  {{Moore}}}\ and\ \bibinfo {author} {\bibfnamefont {L.}~\bibnamefont
  {{Balents}}},\ }\href {\doibase 10.1103/PhysRevB.75.121306} {\bibfield
  {journal} {\bibinfo  {journal} {\prb}\ }\textbf {\bibinfo {volume} {75}},\
  \bibinfo {eid} {121306} (\bibinfo {year} {2007})},\ \Eprint
  {http://arxiv.org/abs/cond-mat/0607314} {cond-mat/0607314} \BibitemShut
  {NoStop}%
\bibitem [{\citenamefont {Mong}\ \emph {et~al.}(2012)\citenamefont {Mong},
  \citenamefont {Bardarson},\ and\ \citenamefont {Moore}}]{MBM12}%
  \BibitemOpen
  \bibfield  {author} {\bibinfo {author} {\bibfnamefont {R.~S.~K.}\
  \bibnamefont {Mong}}, \bibinfo {author} {\bibfnamefont {J.~H.}\ \bibnamefont
  {Bardarson}}, \ and\ \bibinfo {author} {\bibfnamefont {J.~E.}\ \bibnamefont
  {Moore}},\ }\href {\doibase 10.1103/PhysRevLett.108.076804} {\bibfield
  {journal} {\bibinfo  {journal} {Phys. Rev. Lett.}\ }\textbf {\bibinfo
  {volume} {108}},\ \bibinfo {pages} {076804} (\bibinfo {year}
  {2012})}\BibitemShut {NoStop}%
\bibitem [{\citenamefont {Ringel}\ \emph {et~al.}(2012)\citenamefont {Ringel},
  \citenamefont {Kraus},\ and\ \citenamefont {Stern}}]{RKS12}%
  \BibitemOpen
  \bibfield  {author} {\bibinfo {author} {\bibfnamefont {Z.}~\bibnamefont
  {Ringel}}, \bibinfo {author} {\bibfnamefont {Y.~E.}\ \bibnamefont {Kraus}}, \
  and\ \bibinfo {author} {\bibfnamefont {A.}~\bibnamefont {Stern}},\ }\href
  {\doibase 10.1103/PhysRevB.86.045102} {\bibfield  {journal} {\bibinfo
  {journal} {Phys. Rev. B}\ }\textbf {\bibinfo {volume} {86}},\ \bibinfo
  {pages} {045102} (\bibinfo {year} {2012})}\BibitemShut {NoStop}%
\bibitem [{\citenamefont {Yoshimura}\ \emph {et~al.}(2013)\citenamefont
  {Yoshimura}, \citenamefont {Matsumoto}, \citenamefont {Takane},\ and\
  \citenamefont {Imura}}]{YMT13}%
  \BibitemOpen
  \bibfield  {author} {\bibinfo {author} {\bibfnamefont {Y.}~\bibnamefont
  {Yoshimura}}, \bibinfo {author} {\bibfnamefont {A.}~\bibnamefont
  {Matsumoto}}, \bibinfo {author} {\bibfnamefont {Y.}~\bibnamefont {Takane}}, \
  and\ \bibinfo {author} {\bibfnamefont {K.-I.}\ \bibnamefont {Imura}},\ }\href
  {\doibase 10.1103/PhysRevB.88.045408} {\bibfield  {journal} {\bibinfo
  {journal} {Phys. Rev. B}\ }\textbf {\bibinfo {volume} {88}},\ \bibinfo
  {pages} {045408} (\bibinfo {year} {2013})}\BibitemShut {NoStop}%
\bibitem [{\citenamefont {Morimoto}\ and\ \citenamefont
  {Furusaki}(2014)}]{MF14}%
  \BibitemOpen
  \bibfield  {author} {\bibinfo {author} {\bibfnamefont {T.}~\bibnamefont
  {Morimoto}}\ and\ \bibinfo {author} {\bibfnamefont {A.}~\bibnamefont
  {Furusaki}},\ }\href {\doibase 10.1103/PhysRevB.89.035117} {\bibfield
  {journal} {\bibinfo  {journal} {Phys. Rev. B}\ }\textbf {\bibinfo {volume}
  {89}},\ \bibinfo {pages} {035117} (\bibinfo {year} {2014})}\BibitemShut
  {NoStop}%
\bibitem [{\citenamefont {{Ran}}\ \emph {et~al.}(2009)\citenamefont {{Ran}},
  \citenamefont {{Zhang}},\ and\ \citenamefont {{Vishwanath}}}]{Ran}%
  \BibitemOpen
  \bibfield  {author} {\bibinfo {author} {\bibfnamefont {Y.}~\bibnamefont
  {{Ran}}}, \bibinfo {author} {\bibfnamefont {Y.}~\bibnamefont {{Zhang}}}, \
  and\ \bibinfo {author} {\bibfnamefont {A.}~\bibnamefont {{Vishwanath}}},\
  }\href {\doibase 10.1038/nphys1220} {\bibfield  {journal} {\bibinfo
  {journal} {Nature Physics}\ }\textbf {\bibinfo {volume} {5}},\ \bibinfo
  {pages} {298} (\bibinfo {year} {2009})}\BibitemShut {NoStop}%
\bibitem [{\citenamefont {Ran}(2010)}]{Ran10}%
  \BibitemOpen
  \bibfield  {author} {\bibinfo {author} {\bibfnamefont {Y.}~\bibnamefont
  {Ran}},\ }\href@noop {} {\bibfield  {journal} {\bibinfo  {journal} {arXiv
  preprint arXiv:1006.5454}\ } (\bibinfo {year} {2010})}\BibitemShut {NoStop}%
\bibitem [{\citenamefont {Qi}\ and\ \citenamefont {Fu}(2015)}]{Qi2015}%
  \BibitemOpen
  \bibfield  {author} {\bibinfo {author} {\bibfnamefont {Y.}~\bibnamefont
  {Qi}}\ and\ \bibinfo {author} {\bibfnamefont {L.}~\bibnamefont {Fu}},\ }\href
  {\doibase 10.1103/PhysRevLett.115.236801} {\bibfield  {journal} {\bibinfo
  {journal} {Phys. Rev. Lett.}\ }\textbf {\bibinfo {volume} {115}},\ \bibinfo
  {pages} {236801} (\bibinfo {year} {2015})}\BibitemShut {NoStop}%
\bibitem [{\citenamefont {Mross}\ \emph {et~al.}(2016)\citenamefont {Mross},
  \citenamefont {Essin}, \citenamefont {Alicea},\ and\ \citenamefont
  {Stern}}]{Mross2016}%
  \BibitemOpen
  \bibfield  {author} {\bibinfo {author} {\bibfnamefont {D.~F.}\ \bibnamefont
  {Mross}}, \bibinfo {author} {\bibfnamefont {A.}~\bibnamefont {Essin}},
  \bibinfo {author} {\bibfnamefont {J.}~\bibnamefont {Alicea}}, \ and\ \bibinfo
  {author} {\bibfnamefont {A.}~\bibnamefont {Stern}},\ }\href {\doibase
  10.1103/PhysRevLett.116.036803} {\bibfield  {journal} {\bibinfo  {journal}
  {Phys. Rev. Lett.}\ }\textbf {\bibinfo {volume} {116}},\ \bibinfo {pages}
  {036803} (\bibinfo {year} {2016})}\BibitemShut {NoStop}%
\bibitem [{\citenamefont {Rasche}\ \emph {et~al.}(2013)\citenamefont {Rasche},
  \citenamefont {Isaeva}, \citenamefont {Ruck}, \citenamefont {Borisenko},
  \citenamefont {Zabolotnyy}, \citenamefont {B{\"u}chner}, \citenamefont
  {Koepernik}, \citenamefont {Ortix}, \citenamefont {Richter},\ and\
  \citenamefont {van~den Brink}}]{RIR13}%
  \BibitemOpen
  \bibfield  {author} {\bibinfo {author} {\bibfnamefont {B.}~\bibnamefont
  {Rasche}}, \bibinfo {author} {\bibfnamefont {A.}~\bibnamefont {Isaeva}},
  \bibinfo {author} {\bibfnamefont {M.}~\bibnamefont {Ruck}}, \bibinfo {author}
  {\bibfnamefont {S.}~\bibnamefont {Borisenko}}, \bibinfo {author}
  {\bibfnamefont {V.}~\bibnamefont {Zabolotnyy}}, \bibinfo {author}
  {\bibfnamefont {B.}~\bibnamefont {B{\"u}chner}}, \bibinfo {author}
  {\bibfnamefont {K.}~\bibnamefont {Koepernik}}, \bibinfo {author}
  {\bibfnamefont {C.}~\bibnamefont {Ortix}}, \bibinfo {author} {\bibfnamefont
  {M.}~\bibnamefont {Richter}}, \ and\ \bibinfo {author} {\bibfnamefont
  {J.}~\bibnamefont {van~den Brink}},\ }\href@noop {} {\bibfield  {journal}
  {\bibinfo  {journal} {Nat. Mater.}\ }\textbf {\bibinfo {volume} {12}},\
  \bibinfo {pages} {422} (\bibinfo {year} {2013})}\BibitemShut {NoStop}%
\bibitem [{\citenamefont {Yan}\ \emph {et~al.}(2012)\citenamefont {Yan},
  \citenamefont {M\"uchler},\ and\ \citenamefont {Felser}}]{YMF12}%
  \BibitemOpen
  \bibfield  {author} {\bibinfo {author} {\bibfnamefont {B.}~\bibnamefont
  {Yan}}, \bibinfo {author} {\bibfnamefont {L.}~\bibnamefont {M\"uchler}}, \
  and\ \bibinfo {author} {\bibfnamefont {C.}~\bibnamefont {Felser}},\ }\href
  {\doibase 10.1103/PhysRevLett.109.116406} {\bibfield  {journal} {\bibinfo
  {journal} {Phys. Rev. Lett.}\ }\textbf {\bibinfo {volume} {109}},\ \bibinfo
  {pages} {116406} (\bibinfo {year} {2012})}\BibitemShut {NoStop}%
\bibitem [{\citenamefont {Tang}\ \emph {et~al.}(2014)\citenamefont {Tang},
  \citenamefont {Yan}, \citenamefont {Cao}, \citenamefont {Wu}, \citenamefont
  {Felser},\ and\ \citenamefont {Duan}}]{TYC14}%
  \BibitemOpen
  \bibfield  {author} {\bibinfo {author} {\bibfnamefont {P.}~\bibnamefont
  {Tang}}, \bibinfo {author} {\bibfnamefont {B.}~\bibnamefont {Yan}}, \bibinfo
  {author} {\bibfnamefont {W.}~\bibnamefont {Cao}}, \bibinfo {author}
  {\bibfnamefont {S.-C.}\ \bibnamefont {Wu}}, \bibinfo {author} {\bibfnamefont
  {C.}~\bibnamefont {Felser}}, \ and\ \bibinfo {author} {\bibfnamefont
  {W.}~\bibnamefont {Duan}},\ }\href {\doibase 10.1103/PhysRevB.89.041409}
  {\bibfield  {journal} {\bibinfo  {journal} {Phys. Rev. B}\ }\textbf {\bibinfo
  {volume} {89}},\ \bibinfo {pages} {041409} (\bibinfo {year}
  {2014})}\BibitemShut {NoStop}%
\bibitem [{\citenamefont {Yang}\ \emph {et~al.}(2014)\citenamefont {Yang},
  \citenamefont {Liu}, \citenamefont {Fu}, \citenamefont {Duan},\ and\
  \citenamefont {Liu}}]{YLF14}%
  \BibitemOpen
  \bibfield  {author} {\bibinfo {author} {\bibfnamefont {G.}~\bibnamefont
  {Yang}}, \bibinfo {author} {\bibfnamefont {J.}~\bibnamefont {Liu}}, \bibinfo
  {author} {\bibfnamefont {L.}~\bibnamefont {Fu}}, \bibinfo {author}
  {\bibfnamefont {W.}~\bibnamefont {Duan}}, \ and\ \bibinfo {author}
  {\bibfnamefont {C.}~\bibnamefont {Liu}},\ }\href {\doibase
  10.1103/PhysRevB.89.085312} {\bibfield  {journal} {\bibinfo  {journal} {Phys.
  Rev. B}\ }\textbf {\bibinfo {volume} {89}},\ \bibinfo {pages} {085312}
  (\bibinfo {year} {2014})}\BibitemShut {NoStop}%
\bibitem [{\citenamefont {Liu}\ \emph {et~al.}(2016)\citenamefont {Liu},
  \citenamefont {Zhou}, \citenamefont {Yao},\ and\ \citenamefont
  {Zhang}}]{LZY16}%
  \BibitemOpen
  \bibfield  {author} {\bibinfo {author} {\bibfnamefont {C.-C.}\ \bibnamefont
  {Liu}}, \bibinfo {author} {\bibfnamefont {J.-J.}\ \bibnamefont {Zhou}},
  \bibinfo {author} {\bibfnamefont {Y.}~\bibnamefont {Yao}}, \ and\ \bibinfo
  {author} {\bibfnamefont {F.}~\bibnamefont {Zhang}},\ }\href {\doibase
  10.1103/PhysRevLett.116.066801} {\bibfield  {journal} {\bibinfo  {journal}
  {Phys. Rev. Lett.}\ }\textbf {\bibinfo {volume} {116}},\ \bibinfo {pages}
  {066801} (\bibinfo {year} {2016})}\BibitemShut {NoStop}%
\bibitem [{\citenamefont {Fu}\ and\ \citenamefont
  {Kane}(2007)}]{FuKaneInversion}%
  \BibitemOpen
  \bibfield  {author} {\bibinfo {author} {\bibfnamefont {L.}~\bibnamefont
  {Fu}}\ and\ \bibinfo {author} {\bibfnamefont {C.~L.}\ \bibnamefont {Kane}},\
  }\href {\doibase 10.1103/PhysRevB.76.045302} {\bibfield  {journal} {\bibinfo
  {journal} {Phys. Rev. B}\ }\textbf {\bibinfo {volume} {76}},\ \bibinfo
  {pages} {045302} (\bibinfo {year} {2007})}\BibitemShut {NoStop}%
\bibitem [{\citenamefont {Fang}\ \emph {et~al.}(2012)\citenamefont {Fang},
  \citenamefont {Gilbert},\ and\ \citenamefont {Bernevig}}]{BernevigPGS}%
  \BibitemOpen
  \bibfield  {author} {\bibinfo {author} {\bibfnamefont {C.}~\bibnamefont
  {Fang}}, \bibinfo {author} {\bibfnamefont {M.~J.}\ \bibnamefont {Gilbert}}, \
  and\ \bibinfo {author} {\bibfnamefont {B.~A.}\ \bibnamefont {Bernevig}},\
  }\href {\doibase 10.1103/PhysRevB.86.115112} {\bibfield  {journal} {\bibinfo
  {journal} {Phys. Rev. B}\ }\textbf {\bibinfo {volume} {86}},\ \bibinfo
  {pages} {115112} (\bibinfo {year} {2012})}\BibitemShut {NoStop}%
\bibitem [{\citenamefont {Benalcazar}\ \emph {et~al.}(2014)\citenamefont
  {Benalcazar}, \citenamefont {Teo},\ and\ \citenamefont {Hughes}}]{HughesPGS}%
  \BibitemOpen
  \bibfield  {author} {\bibinfo {author} {\bibfnamefont {W.~A.}\ \bibnamefont
  {Benalcazar}}, \bibinfo {author} {\bibfnamefont {J.~C.~Y.}\ \bibnamefont
  {Teo}}, \ and\ \bibinfo {author} {\bibfnamefont {T.~L.}\ \bibnamefont
  {Hughes}},\ }\href {\doibase 10.1103/PhysRevB.89.224503} {\bibfield
  {journal} {\bibinfo  {journal} {Phys. Rev. B}\ }\textbf {\bibinfo {volume}
  {89}},\ \bibinfo {pages} {224503} (\bibinfo {year} {2014})}\BibitemShut
  {NoStop}%
\bibitem [{\citenamefont {{Slager}}\ \emph {et~al.}(2013)\citenamefont
  {{Slager}}, \citenamefont {{Mesaros}}, \citenamefont {{Juri{\v c}i{\'c}}},\
  and\ \citenamefont {{Zaanen}}}]{Zaanen}%
  \BibitemOpen
  \bibfield  {author} {\bibinfo {author} {\bibfnamefont {R.-J.}\ \bibnamefont
  {{Slager}}}, \bibinfo {author} {\bibfnamefont {A.}~\bibnamefont {{Mesaros}}},
  \bibinfo {author} {\bibfnamefont {V.}~\bibnamefont {{Juri{\v c}i{\'c}}}}, \
  and\ \bibinfo {author} {\bibfnamefont {J.}~\bibnamefont {{Zaanen}}},\ }\href
  {\doibase 10.1038/nphys2513} {\bibfield  {journal} {\bibinfo  {journal}
  {Nature Physics}\ }\textbf {\bibinfo {volume} {9}},\ \bibinfo {pages} {98}
  (\bibinfo {year} {2013})},\ \Eprint {http://arxiv.org/abs/1209.2610}
  {arXiv:1209.2610 [cond-mat.mes-hall]} \BibitemShut {NoStop}%
\bibitem [{\citenamefont {{Chiu}}\ \emph {et~al.}(2013)\citenamefont {{Chiu}},
  \citenamefont {{Yao}},\ and\ \citenamefont {{Ryu}}}]{Chiu}%
  \BibitemOpen
  \bibfield  {author} {\bibinfo {author} {\bibfnamefont {C.-K.}\ \bibnamefont
  {{Chiu}}}, \bibinfo {author} {\bibfnamefont {H.}~\bibnamefont {{Yao}}}, \
  and\ \bibinfo {author} {\bibfnamefont {S.}~\bibnamefont {{Ryu}}},\ }\href
  {\doibase 10.1103/PhysRevB.88.075142} {\bibfield  {journal} {\bibinfo
  {journal} {\prb}\ }\textbf {\bibinfo {volume} {88}},\ \bibinfo {eid} {075142}
  (\bibinfo {year} {2013})},\ \Eprint {http://arxiv.org/abs/1303.1843}
  {arXiv:1303.1843 [cond-mat.mes-hall]} \BibitemShut {NoStop}%
\bibitem [{\citenamefont {{Morimoto}}\ and\ \citenamefont
  {{Furusaki}}(2013)}]{Morimoto}%
  \BibitemOpen
  \bibfield  {author} {\bibinfo {author} {\bibfnamefont {T.}~\bibnamefont
  {{Morimoto}}}\ and\ \bibinfo {author} {\bibfnamefont {A.}~\bibnamefont
  {{Furusaki}}},\ }\href {\doibase 10.1103/PhysRevB.88.125129} {\bibfield
  {journal} {\bibinfo  {journal} {\prb}\ }\textbf {\bibinfo {volume} {88}},\
  \bibinfo {eid} {125129} (\bibinfo {year} {2013})},\ \Eprint
  {http://arxiv.org/abs/1306.2505} {arXiv:1306.2505 [cond-mat.mes-hall]}
  \BibitemShut {NoStop}%
\bibitem [{\citenamefont {{Lu}}\ and\ \citenamefont {{Lee}}(2014)}]{Lu}%
  \BibitemOpen
  \bibfield  {author} {\bibinfo {author} {\bibfnamefont {Y.-M.}\ \bibnamefont
  {{Lu}}}\ and\ \bibinfo {author} {\bibfnamefont {D.-H.}\ \bibnamefont
  {{Lee}}},\ }\href@noop {} {\bibfield  {journal} {\bibinfo  {journal} {ArXiv
  e-prints}\ } (\bibinfo {year} {2014})},\ \Eprint
  {http://arxiv.org/abs/1403.5558} {arXiv:1403.5558 [cond-mat.mes-hall]}
  \BibitemShut {NoStop}%
\bibitem [{\citenamefont {{Fang}}\ and\ \citenamefont {{Fu}}(2015)}]{FuGlide}%
  \BibitemOpen
  \bibfield  {author} {\bibinfo {author} {\bibfnamefont {C.}~\bibnamefont
  {{Fang}}}\ and\ \bibinfo {author} {\bibfnamefont {L.}~\bibnamefont {{Fu}}},\
  }\href {\doibase 10.1103/PhysRevB.91.161105} {\bibfield  {journal} {\bibinfo
  {journal} {\prb}\ }\textbf {\bibinfo {volume} {91}},\ \bibinfo {eid} {161105}
  (\bibinfo {year} {2015})}\BibitemShut {NoStop}%
\bibitem [{\citenamefont {{Shiozaki}}\ and\ \citenamefont
  {{Sato}}(2014)}]{Shiozaki}%
  \BibitemOpen
  \bibfield  {author} {\bibinfo {author} {\bibfnamefont {K.}~\bibnamefont
  {{Shiozaki}}}\ and\ \bibinfo {author} {\bibfnamefont {M.}~\bibnamefont
  {{Sato}}},\ }\href {\doibase 10.1103/PhysRevB.90.165114} {\bibfield
  {journal} {\bibinfo  {journal} {\prb}\ }\textbf {\bibinfo {volume} {90}},\
  \bibinfo {eid} {165114} (\bibinfo {year} {2014})},\ \Eprint
  {http://arxiv.org/abs/1403.3331} {arXiv:1403.3331 [cond-mat.mes-hall]}
  \BibitemShut {NoStop}%
\bibitem [{\citenamefont {{Mong}}\ \emph {et~al.}(2010)\citenamefont {{Mong}},
  \citenamefont {{Essin}},\ and\ \citenamefont {{Moore}}}]{Mong}%
  \BibitemOpen
  \bibfield  {author} {\bibinfo {author} {\bibfnamefont {R.~S.~K.}\
  \bibnamefont {{Mong}}}, \bibinfo {author} {\bibfnamefont {A.~M.}\
  \bibnamefont {{Essin}}}, \ and\ \bibinfo {author} {\bibfnamefont {J.~E.}\
  \bibnamefont {{Moore}}},\ }\href {\doibase 10.1103/PhysRevB.81.245209}
  {\bibfield  {journal} {\bibinfo  {journal} {\prb}\ }\textbf {\bibinfo
  {volume} {81}},\ \bibinfo {eid} {245209} (\bibinfo {year} {2010})},\ \Eprint
  {http://arxiv.org/abs/1004.1403} {arXiv:1004.1403 [cond-mat.mes-hall]}
  \BibitemShut {NoStop}%
\bibitem [{\citenamefont {{Liu}}\ \emph {et~al.}(2014)\citenamefont {{Liu}},
  \citenamefont {{Zhang}},\ and\ \citenamefont {{VanLeeuwen}}}]{LiuNS}%
  \BibitemOpen
  \bibfield  {author} {\bibinfo {author} {\bibfnamefont {C.-X.}\ \bibnamefont
  {{Liu}}}, \bibinfo {author} {\bibfnamefont {R.-X.}\ \bibnamefont {{Zhang}}},
  \ and\ \bibinfo {author} {\bibfnamefont {B.~K.}\ \bibnamefont
  {{VanLeeuwen}}},\ }\href {\doibase 10.1103/PhysRevB.90.085304} {\bibfield
  {journal} {\bibinfo  {journal} {\prb}\ }\textbf {\bibinfo {volume} {90}},\
  \bibinfo {eid} {085304} (\bibinfo {year} {2014})},\ \Eprint
  {http://arxiv.org/abs/1308.4717} {arXiv:1308.4717 [cond-mat.mes-hall]}
  \BibitemShut {NoStop}%
\bibitem [{\citenamefont {Thouless}\ \emph {et~al.}(1982)\citenamefont
  {Thouless}, \citenamefont {Kohmoto}, \citenamefont {Nightingale},\ and\
  \citenamefont {den Nijs}}]{Thouless1982}%
  \BibitemOpen
  \bibfield  {author} {\bibinfo {author} {\bibfnamefont {D.~J.}\ \bibnamefont
  {Thouless}}, \bibinfo {author} {\bibfnamefont {M.}~\bibnamefont {Kohmoto}},
  \bibinfo {author} {\bibfnamefont {M.~P.}\ \bibnamefont {Nightingale}}, \ and\
  \bibinfo {author} {\bibfnamefont {M.}~\bibnamefont {den Nijs}},\ }\href
  {http://link.aps.org/doi/10.1103/PhysRevLett.49.405} {\bibfield  {journal}
  {\bibinfo  {journal} {Phys. Rev. Lett.}\ }\textbf {\bibinfo {volume} {49}},\
  \bibinfo {pages} {405} (\bibinfo {year} {1982})}\BibitemShut {NoStop}%
\bibitem [{\citenamefont {Nagaosa}\ \emph {et~al.}(2010)\citenamefont
  {Nagaosa}, \citenamefont {Sinova}, \citenamefont {Onoda}, \citenamefont
  {MacDonald},\ and\ \citenamefont {Ong}}]{Nagaosa2010}%
  \BibitemOpen
  \bibfield  {author} {\bibinfo {author} {\bibfnamefont {N.}~\bibnamefont
  {Nagaosa}}, \bibinfo {author} {\bibfnamefont {J.}~\bibnamefont {Sinova}},
  \bibinfo {author} {\bibfnamefont {S.}~\bibnamefont {Onoda}}, \bibinfo
  {author} {\bibfnamefont {A.~H.}\ \bibnamefont {MacDonald}}, \ and\ \bibinfo
  {author} {\bibfnamefont {N.~P.}\ \bibnamefont {Ong}},\ }\href
  {http://link.aps.org/doi/10.1103/RevModPhys.82.1539} {\bibfield  {journal}
  {\bibinfo  {journal} {Rev. Mod. Phys.}\ }\textbf {\bibinfo {volume} {82}},\
  \bibinfo {pages} {1539} (\bibinfo {year} {2010})}\BibitemShut {NoStop}%
\bibitem [{\citenamefont {Halperin}(1987)}]{Halperin1987}%
  \BibitemOpen
  \bibfield  {author} {\bibinfo {author} {\bibfnamefont {B.~I.}\ \bibnamefont
  {Halperin}},\ }\href {http://iopscience.iop.org/1347-4065/26/S3-3/1913}
  {\bibfield  {journal} {\bibinfo  {journal} {Japanese Journal of Applied
  Physics}\ }\textbf {\bibinfo {volume} {26}},\ \bibinfo {pages} {1913}
  (\bibinfo {year} {1987})}\BibitemShut {NoStop}%
\bibitem [{\citenamefont {{Parameswaran}}\ \emph {et~al.}(2013)\citenamefont
  {{Parameswaran}}, \citenamefont {{Turner}}, \citenamefont {{Arovas}},\ and\
  \citenamefont {{Vishwanath}}}]{SidNS}%
  \BibitemOpen
  \bibfield  {author} {\bibinfo {author} {\bibfnamefont {S.~A.}\ \bibnamefont
  {{Parameswaran}}}, \bibinfo {author} {\bibfnamefont {A.~M.}\ \bibnamefont
  {{Turner}}}, \bibinfo {author} {\bibfnamefont {D.~P.}\ \bibnamefont
  {{Arovas}}}, \ and\ \bibinfo {author} {\bibfnamefont {A.}~\bibnamefont
  {{Vishwanath}}},\ }\href {\doibase 10.1038/nphys2600} {\bibfield  {journal}
  {\bibinfo  {journal} {Nature Physics}\ }\textbf {\bibinfo {volume} {9}},\
  \bibinfo {pages} {299} (\bibinfo {year} {2013})},\ \Eprint
  {http://arxiv.org/abs/1212.0557} {arXiv:1212.0557 [cond-mat.str-el]}
  \BibitemShut {NoStop}%
\bibitem [{\citenamefont {{Watanabe}}\ \emph {et~al.}(2015)\citenamefont
  {{Watanabe}}, \citenamefont {{Po}}, \citenamefont {{Vishwanath}},\ and\
  \citenamefont {{Zaletel}}}]{WatanabeNS}%
  \BibitemOpen
  \bibfield  {author} {\bibinfo {author} {\bibfnamefont {H.}~\bibnamefont
  {{Watanabe}}}, \bibinfo {author} {\bibfnamefont {H.~C.}\ \bibnamefont
  {{Po}}}, \bibinfo {author} {\bibfnamefont {A.}~\bibnamefont {{Vishwanath}}},
  \ and\ \bibinfo {author} {\bibfnamefont {M.~P.}\ \bibnamefont {{Zaletel}}},\
  }\href@noop {} {\bibfield  {journal} {\bibinfo  {journal} {ArXiv e-prints}\ }
  (\bibinfo {year} {2015})},\ \Eprint {http://arxiv.org/abs/1505.04193}
  {arXiv:1505.04193 [cond-mat.str-el]} \BibitemShut {NoStop}%
\bibitem [{\citenamefont {Streda}(1982)}]{Streda1982}%
  \BibitemOpen
  \bibfield  {author} {\bibinfo {author} {\bibfnamefont {P.}~\bibnamefont
  {Streda}},\ }\href {http://stacks.iop.org/0022-3719/15/i=36/a=006} {\bibfield
   {journal} {\bibinfo  {journal} {Journal of Physics C: Solid State Physics}\
  }\textbf {\bibinfo {volume} {15}},\ \bibinfo {pages} {L1299} (\bibinfo {year}
  {1982})}\BibitemShut {NoStop}%
\bibitem [{\citenamefont {Fu}\ \emph {et~al.}(2007)\citenamefont {Fu},
  \citenamefont {Kane},\ and\ \citenamefont {Mele}}]{FuKane2007}%
  \BibitemOpen
  \bibfield  {author} {\bibinfo {author} {\bibfnamefont {L.}~\bibnamefont
  {Fu}}, \bibinfo {author} {\bibfnamefont {C.~L.}\ \bibnamefont {Kane}}, \ and\
  \bibinfo {author} {\bibfnamefont {E.~J.}\ \bibnamefont {Mele}},\ }\href
  {\doibase 10.1103/PhysRevLett.98.106803} {\bibfield  {journal} {\bibinfo
  {journal} {Phys. Rev. Lett.}\ }\textbf {\bibinfo {volume} {98}},\ \bibinfo
  {pages} {106803} (\bibinfo {year} {2007})}\BibitemShut {NoStop}%
\bibitem [{Note1()}]{Note1}%
  \BibitemOpen
  \bibinfo {note} {A priori the constraints depend on the \protect \emph
  {arithmetic crystal class} that is determined by the point group (geometric
  crystal class) and its action on the translations (Bravais lattice). There
  are 73 arythmetic crystal classes in 3d in one-to-one correspondence to
  symmorphic space groups. The reciprocal space group has the same point group
  as the space group, but the reciprocal Bravais lattice may have different
  centering. Direct enumeration\cite {ITA,Aroyo2006,Aroyo2014} reveals that the
  result in fact only depends on the Bravais-lattice, this can be rationalized
  noting that a Bravais-lattice contains arithmetic crystal classes that have a
  large enough point group to force certain relations among lattice vectors by
  mapping them to each other. As $\protect \mathbf {G}^{\nu }$ takes values on
  the same reciprocal Bravais-lattice with half the spacing, the constraints
  are the same for every arithmetic crystal class in a
  Bravais-lattice.}\BibitemShut {Stop}%
\bibitem [{\citenamefont {{Wondratschek}}\ and\ \citenamefont
  {{M{\"u}ller}}(2008)}]{ITA}%
  \BibitemOpen
  \bibfield  {author} {\bibinfo {author} {\bibfnamefont {H.}~\bibnamefont
  {{Wondratschek}}}\ and\ \bibinfo {author} {\bibfnamefont {U.}~\bibnamefont
  {{M{\"u}ller}}},\ }\href@noop {} {\emph {\bibinfo {title} {International
  Tables for Crystallography,Volume A1, Edited by H.~Wondratschek, and
  U.~M{\"u}ller.~Berlin: Springer, 2008.~ISBN: 978-1-4020-8208-5}}}\ (\bibinfo
  {year} {2008})\BibitemShut {NoStop}%
\bibitem [{\citenamefont {{Aroyo}}\ \emph {et~al.}(2006)\citenamefont
  {{Aroyo}}, \citenamefont {{Perez-Mato}}, \citenamefont {{Capillas}},
  \citenamefont {{Kroumova}}, \citenamefont {{Ivantchev}}, \citenamefont
  {{Madariaga}}, \citenamefont {{Kirov}},\ and\ \citenamefont
  {{Wondratschek}}}]{Aroyo2006}%
  \BibitemOpen
  \bibfield  {author} {\bibinfo {author} {\bibfnamefont {M.~I.}\ \bibnamefont
  {{Aroyo}}}, \bibinfo {author} {\bibfnamefont {J.~M.}\ \bibnamefont
  {{Perez-Mato}}}, \bibinfo {author} {\bibfnamefont {C.}~\bibnamefont
  {{Capillas}}}, \bibinfo {author} {\bibfnamefont {E.}~\bibnamefont
  {{Kroumova}}}, \bibinfo {author} {\bibfnamefont {S.}~\bibnamefont
  {{Ivantchev}}}, \bibinfo {author} {\bibfnamefont {G.}~\bibnamefont
  {{Madariaga}}}, \bibinfo {author} {\bibfnamefont {A.}~\bibnamefont
  {{Kirov}}}, \ and\ \bibinfo {author} {\bibfnamefont {H.}~\bibnamefont
  {{Wondratschek}}},\ }\href {\doibase 10.1524/zkri.2006.221.1.15} {\bibfield
  {journal} {\bibinfo  {journal} {Zeitschrift fur Kristallographie}\ }\textbf
  {\bibinfo {volume} {221}},\ \bibinfo {pages} {15} (\bibinfo {year}
  {2006})}\BibitemShut {NoStop}%
\bibitem [{\citenamefont {Aroyo}\ \emph {et~al.}(2014)\citenamefont {Aroyo},
  \citenamefont {Orobengoa}, \citenamefont {de~la Flor}, \citenamefont {Tasci},
  \citenamefont {Perez-Mato},\ and\ \citenamefont {Wondratschek}}]{Aroyo2014}%
  \BibitemOpen
  \bibfield  {author} {\bibinfo {author} {\bibfnamefont {M.~I.}\ \bibnamefont
  {Aroyo}}, \bibinfo {author} {\bibfnamefont {D.}~\bibnamefont {Orobengoa}},
  \bibinfo {author} {\bibfnamefont {G.}~\bibnamefont {de~la Flor}}, \bibinfo
  {author} {\bibfnamefont {E.~S.}\ \bibnamefont {Tasci}}, \bibinfo {author}
  {\bibfnamefont {J.~M.}\ \bibnamefont {Perez-Mato}}, \ and\ \bibinfo {author}
  {\bibfnamefont {H.}~\bibnamefont {Wondratschek}},\ }\href {\doibase
  10.1107/S205327331303091X} {\bibfield  {journal} {\bibinfo  {journal} {Acta
  Crystallographica Section A}\ }\textbf {\bibinfo {volume} {70}},\ \bibinfo
  {pages} {126} (\bibinfo {year} {2014})}\BibitemShut {NoStop}%
\bibitem [{\citenamefont {Varjas}\ \emph {et~al.}(2015)\citenamefont {Varjas},
  \citenamefont {de~Juan},\ and\ \citenamefont {Lu}}]{Varjas2015}%
  \BibitemOpen
  \bibfield  {author} {\bibinfo {author} {\bibfnamefont {D.}~\bibnamefont
  {Varjas}}, \bibinfo {author} {\bibfnamefont {F.}~\bibnamefont {de~Juan}}, \
  and\ \bibinfo {author} {\bibfnamefont {Y.-M.}\ \bibnamefont {Lu}},\ }\href
  {\doibase 10.1103/PhysRevB.92.195116} {\bibfield  {journal} {\bibinfo
  {journal} {Phys. Rev. B}\ }\textbf {\bibinfo {volume} {92}},\ \bibinfo
  {pages} {195116} (\bibinfo {year} {2015})}\BibitemShut {NoStop}%
\bibitem [{\citenamefont {Alexandradinata}\ \emph {et~al.}(2014)\citenamefont
  {Alexandradinata}, \citenamefont {Dai},\ and\ \citenamefont
  {Bernevig}}]{Alexandradinata2014}%
  \BibitemOpen
  \bibfield  {author} {\bibinfo {author} {\bibfnamefont {A.}~\bibnamefont
  {Alexandradinata}}, \bibinfo {author} {\bibfnamefont {X.}~\bibnamefont
  {Dai}}, \ and\ \bibinfo {author} {\bibfnamefont {B.~A.}\ \bibnamefont
  {Bernevig}},\ }\href {\doibase 10.1103/PhysRevB.89.155114} {\bibfield
  {journal} {\bibinfo  {journal} {Phys. Rev. B}\ }\textbf {\bibinfo {volume}
  {89}},\ \bibinfo {pages} {155114} (\bibinfo {year} {2014})}\BibitemShut
  {NoStop}%
\bibitem [{\citenamefont {de~Juan}\ \emph {et~al.}(2014)\citenamefont
  {de~Juan}, \citenamefont {R\"uegg},\ and\ \citenamefont {Lee}}]{deJuan2014}%
  \BibitemOpen
  \bibfield  {author} {\bibinfo {author} {\bibfnamefont {F.}~\bibnamefont
  {de~Juan}}, \bibinfo {author} {\bibfnamefont {A.}~\bibnamefont {R\"uegg}}, \
  and\ \bibinfo {author} {\bibfnamefont {D.-H.}\ \bibnamefont {Lee}},\ }\href
  {\doibase 10.1103/PhysRevB.89.161117} {\bibfield  {journal} {\bibinfo
  {journal} {Phys. Rev. B}\ }\textbf {\bibinfo {volume} {89}},\ \bibinfo
  {pages} {161117} (\bibinfo {year} {2014})}\BibitemShut {NoStop}%
\end{thebibliography}%

\appendix

\section{Conventions for Bloch functions}
\label{sec:conv}

There are two widely used conventions to define the Bloch basis functions. When appropriate we use the convention where we define Bloch basis functions $\ket{\widetilde{\chi}_{\bk}^a}$ in terms of the orbitals of the unit cell $\ket{\widetilde{\chi}_{\bk}^a} = \sum_{\bR} e^{i \bk \bR} \ket{\phi_{\bR}^a}$ where $\bR$ is the unit cell coordinate and $a$ the orbital index. Note the absence of phase factors corresponding to the position of the orbitals within the unit cell, so the basis functions are strictly periodic in the BZ. While in this convention the information about the position of the orbitals is lost, thus the polarizations computed via Berry vector potential integrals do not equal the true Wannier center positions, the Bloch Hamiltonian is BZ periodic, making some derivations more transparent.

In the other convention we define $\ket{\chi_{\bk}^a} = \sum_{\bR} e^{i \bk (\bR+\mathbf{r}_a)} \ket{\phi_{\bR}^a}$ where $\mathbf{r}_a$ is the position of the $a$-th orbital in the unit cell. The two conventions are related by the operator $W_{\bk}$ with $W^{ab}_{\bk} = \delta^{ab} e^{-i \bk \mathbf{r}_a}$ such that $\ket{\chi_{\bk}^a} = \pars{W_{\bk}^{-1}}^{ab} \ket{\widetilde{\chi}_{\bk}^b}$ so the coefficients of of Bloch wavefunctions transform as $n_{\bk}^a = W_{\bk}^{ab} \widetilde{n}_{\bk}^{ab}$. Consequently, operators expanded in this basis (including the Bloch Hamiltonian) satisfy $O_{\bk+\bG} = W_{\bG} O_{\bk} W_{\bG}^{-1}$ where $\bG$ is a primitive reciprocal lattice vector. $W_{\bG}: \mathcal{H}_{\bk}\to\mathcal{H}_{\bk+\bG}$ is acting between the Hilbert spaces of the \emph{coefficients} of the wavefunctions in this basis. This convention, using the coefficients only (e.g. $\cA^{nm} = i n\dag \dee m$), is usually assumed in formulae for electromagnetic response, as the naive Peierls substitution $\bk \to \bk + \mathbf{A}$ only gives the correct phase factor for hopping in this case. The two conventions give equivalent results for quantized topological indices in most symmorphic cases, provided there is a continuous, symmetry preserving deformation of the lattice, such that all the orbitals are brought to the same point in the unit cell. In nonsymmorphic lattices however, this is never possible, as the shortest orbit of a point in the unit cell under the symmetry group modulo lattice vectors is longer than one, there is no crystal with one site per unit cell obeying a nonsymmorphic symmetry. For example with an $n$-fold screw translating in the $z$ direction one needs at least $n$ lattice sites that can be arranged such that the positions are $\mathbf{r}_a = \mathbf{a}_z a/ n$ for $a = 1 \textellipsis n$, so $W_{\bG_z}^{ab} = \delta^{ab} e^{2\pi i a/n}$ and $W_{\bG} = \id$ for perpendicular directions.

We remark that in both bases global antiunitary transformations, such as $\cT = \cK \sigma_y$ act as constant operators in $\bk$-space. To switch conventions, one must transform them same as other operators, $\widetilde{\cT} = W_{-\bk} \cT W_{\bk}^{-1} = \cT$, where we used that $W_{\bk}^{-1} = W_{-\bk} = W_{\bk}^*$ as $W$ is diagonal and proportional to the identity in spin space.

\section{Proof for transformation properties of weak indices}
\label{sec:trfproof}

First we review the representations of space group operations in $\bk$-space. We use the convention (Appendix~\ref{sec:conv}) with Bloch basis functions $\ket{\chi_{\bk}^{\bx l}} = \sum_{\bR} e^{i \bk (\bR+\bx)} \ket{\phi_{\bR+\bx}^{l}}$, where we split the orbital index $a = (\bx,l)$, $\bx$ labels the sites of the unit cell by their real space position and $l$ is an on-site orbital index accounting for spin, orbital angular momentum, etc. (the values $l$ can take may depend on $\bx$). A useful property of this basis is that it is periodic in the real space coordinate, i.e. $\ket{\chi_{\bk}^{\pars{\bx+\bR} l}} = \ket{\chi_{\bk}^{\bx l}}$ for any lattice vector $\bR$. We emphasize that our treatment is not specific to tight-binding models, the same can be told in the continuum, there $\bx$ is the continuous index for position in the unit cell and $l$ stands for the spin only. To go to the tight-binding approximation, we restrict the Hilbert-space to a finite set of orbitals per unit cell, the only assumption we make is that orbitals centered on different sites span orthogonal subspaces.

Consider a general space-group operation $g = \brac{O \middle| \bt}$ acting on one of the basis states
\eqn{
g \ket{\phi_{\bR + \bx}^{l}} = U_{\bx}^{ll'} \ket{\phi_{g\pars{\bR + \bx}}^{l'}} =
U_{\bx}^{ll'} \ket{\phi_{O\pars{\bR + \bx}+\bt}^{l'}}
}
where $U$ is the site and $g$-dependent unitary representation on the local orbitals, a double representation if the model is spinful. Applying this to the Bloch basis functions, with simple algebra we find
\eqn{
g \ket{\chi_{\bk}^{\bx l}} = e^{-i\pars{g\bk}\bt} U_{\bx}^{ll'} \ket{\chi_{g\bk}^{g\bx, l'}}
}
with $g\bk = O\bk$ and $g\bx = O\bx + \bt$ that is understood as a permutation of sites at the same Wyckoff position. Grouping indices back together, this can be written as $g \ket{\chi_{\bk}^{a}} = e^{-i\pars{g\bk}\bt} U^{ba} \ket{\chi_{g\bk}^{b}}$

The key observation is that in this basis the $\bk$-dependence decouples as a single factor proportional to the identity. Consider the transformation of a Bloch eigenstate in the $n$-th band $\ket{n_{\bk}} = n^a_{\bk} \ket{\chi_{\bk}^{a}}$. The symmetry transformation results in a state at $g\bk$, the coefficients transform as $(gn)_{g\bk}^a = e^{-i(g\bk)\bt} U^{ab} n_{\bk}^b$ or in a compact notation $g \pars{n_{\bk}} = (gn)_{g\bk} = e^{-i\pars{g\bk}\bt} U n_{\bk}$. As $g$ is a symmetry operation, the transformed state is again an eigenstate of the Bloch Hamiltonian with the same energy, but at $g\bk$. As a consequence, the transformation of occupied band projector operator $\cP_{\bk} = \sum_{n\in\textnormal{occ.}} n_{\bk} n_{\bk}\dag$ reads
\eqn{
(g\cP)_{g\bk} = \sum_{n\in\textnormal{occ.}} (gn)_{g\bk}(gn)_{g\bk}\dag = U \cP_{\bk} U\dag.
}
So if $g$ is a symmetry, such that $(g\cP)_{\bk} = \cP_{\bk}$, any gauge invariant quantity that can be expressed through $\cP_{\bk}$ is invariant if the $\bk$-space coordinates are transformed accordingly. Examples include\cite{Varjas2015} the Berry curvature $\cF = i \cP \dee \cP \wedge \dee \cP \cP$ and closed loop integrals of the Berry connection $\cA$ (see below).

The ``Hall vector'' as defined in Eqn.~(\ref{eqn:hall conductivity tensor}) may be cast in a coordinate free form, as
\eqn{
\mathbf{\Sigma} = \frac{1}{2\pi} \int \Tr \cF \wedge \dee\bk.
}
To see that it is equal to $\sum_{i=1}^3\bG_i\mathcal{C}_i$, it is sufficient to check that $\mathbf{a}_i\cdot\mathbf{\Sigma}$ is the same in the two cases for all lattice vectors. Simple substitution shows that this vector transforms as an axial vector under all SG operations (i.e. even under inversion, $\mathbf{\Sigma}\to\pars{\det O} O \mathbf{\Sigma}$) as stated in the main text.

For completeness we derive the transformation properties of $\cA$ and $\cF$ under the basis change corresponding to switching between conventions and show that invariants calculated in either convention give the same result. We feel this is necessary because, while the Berry connection for the Bloch basis $\ket{\widetilde{\chi}_{\bk}^a}$ vanishes and one can safely use the coefficients, for the basis $\ket{\chi_{\bk}^a}$ it is nonzero, $\cA_{\chi}^{ab}=i\bra{\chi_{\bk}^a}\dee\ket{\chi_{\bk}^b} = i \delta^{ab} e^{i \bk \bt_a} \bt_a \dee\bk$. This means that one may worry that the formulae in terms of the components in this basis may be missing some terms coming from the derivatives of the basis vectors.

For generality, we consider a transformation $U_{\bk}$ acting on the coefficients, it may either be a basis transformation, or a physical one and let $n'_{\bk} = U_{\bk} n_{\bk}$. We find
\begin{align}
\Tr_{occ.} \cA' &= \sum_n i n^{\prime \dagger} \dee n'  =  \Tr_{occ.} \cA + i \Tr\pars{\cP U\dag \dee U}\label{eqn:intA}\\
\Tr_{occ.} \cF' &= \Tr_{occ.} \dee\cA' = \Tr_{occ.} \cF + i \dee \Tr\pars{\cP U\dag \dee U}
\end{align}
where $\Tr_{occ.}(.) = \sum_n (.)^{nn}$ is the trace over occupied bands, while $\Tr(.) = \sum_a (.)^{aa}$ is the trace over the entire Hilbert-space of the unit cell. We see that as long as $\cP U\dag \dee U$ is unit cell periodic, which is the case for the basis transformation $U_{\bk}^{ab} = W_{\bk}^{ab} = \delta^{ab} e^{i \bk \br_a}$ if $\cP=\widetilde{\cP}$ (BZ periodic convention). The change in $\Tr_{occ.} \int_S \cF$ is fully compensated by the change in $\Tr_{occ.} \int_{\partial S} \cA$ in the formula for the $\bbZ_2$ invariants and vanishes for Chern-numbers.

We note that the expression for $\Tr_{occ.}\oint\cA$ along a non-contractible loop in terms of the projector is modified in the $\ket{\chi}$ basis,
\eqn{
\Tr_{occ.}\oint\cA = i \log \det_+ \pars{W_{-\bG}\prod_{\bk=\bk_0}^{\bk_0-\bG} \cP_{\bk}}
}
where $\det_+$ is the pseudo-determinant of the matrix, which is defined as the product of all nonzero eigenvalues. This is equivalent to calculating the determinant of the restriction to the local occupied space at $\bk_0$, i.e. we evaluate $-i \log\det_+$ as the sum of the complex phases of the nonzero eigenvalues. The reason $W_{\bG}$ appears is the mismatch of the basis at $\bk_0$ and $\bk_0+\bG$.

\section{Wilson Loop Invariants}
\label{sec:WL}
Here we review symmetry properties of Wilson loop (WL) eigenvalues. As a natural consequence of our formalism we find constraints SG symmetry imposes on WL eigenvalues and resulting new topological invariants in topological crystalline insulators. Similar results were obtained in Refs~\onlinecite{BernevigPGS,Alexandradinata2014}, those we generalize for all SG symmetries.

We define Wilson line operators as maps from the unit cell Hilbert space at the start of line $\mathcal{C}$ at $\bk_0$ to the end at $\bk_1$, $\mathcal{W}_{\mathcal{C}}: \mathcal{H}_{\bk_0}\to\mathcal{H}_{\bk_1}$. We use the definition in terms of an ordered discretization of $\mathcal{C}$, $\brac{\bk_i}_{i=1}^N$ with $\bk_1 = \bk _0$ and $\bk_N = \bk_1$:
\eqn{
\mathcal{W}_{\mathcal{C}} = \lim_{N\to\infty}\prod_{i=N}^{1} \cP_{\bk_i}.
}
We use the projector operators $\cP$ expanded in the $\ket{\chi}$ convention, as in this convention it contains all the information about the positions of the sites in the unit cell and is invariant under the choice of the unit cell besides some other useful properties we will prove later. Note that $\mathcal{W}_\mathcal{C}$ only takes nonzero values on the occupied subspace at $\bk_0$ and its image is the occupied space at $\bk_1$. Its restriction to the local occupied spaces is unitary, this is seen by checking that $\mathcal{W}_\mathcal{C}\dag\mathcal{W}_\mathcal{C} = \mathcal{W}_{\mathcal{C}^{-1}}\mathcal{W}_{\mathcal{C}} = \cP_{\bk_0} + O\pars{dk}$ with $dk\to 0$ the typical distance between the discretized $\bk$ points. This means that all of its eigenvalues are either zero or a unit magnitude complex number, with the number of nonzero eigenvalues equal to the number of occupied bands. Moreover, Wilson loop operators restricted to the local occupied spaces form a unitary representation of the groupoid of curves under concatenation as $\mathcal{W}_{\mathcal{C}_2}\mathcal{W}_{\mathcal{C}_1} = \mathcal{W}_{\mathcal{C}_1+\mathcal{C}_2}$ where the end point of $\cC_1$ is the same as the starting point of $\cC_2$.

This definition is consistent with the definition in terms of the Berry connection as
\eqn{
\sum_{n,m\in\textnormal{occ.}}\pars{P\exp\pars{i \int_{\mathcal{C}} \cA}}^{nm} n_{\bk_0} m_{\bk_1}\dag = \mathcal{W}_{\mathcal{C}}
}
where $n_{\bk}$ is the coefficients of an orthonormal set of occupied eigenstates in the $\ket{\chi}$ convention at $\bk$\cite{BernevigPGS,Alexandradinata2014}. In general $\mathcal{W}_{\mathcal{C}}$ is gauge-dependent, however, as we show in the following, for a closed loop its spectrum $\sigma\pars{\mathcal{W}_{\mathcal{C}}}$ (with multiplicities) is well defined.

For a closed loop $\mathcal{C}$ we define
\eqn{
\mathcal{W}_{\mathcal{C}} = W_{-\bG} \pars{\prod_{\bk = \bk_0}^{\bk_0-\bG } \cP_{\bk}}
\label{eqn:closedWL}
}
where $\bG$ is a nonzero reciprocal lattice vector if the loop is non-contractible and the overall displacement along it is $\bG$. We use this definition to make $\mathcal{W}_{\mathcal{C}}$ a mapping from $\mathcal{H}_{\bk_0}$ onto itself, as in our convention the Hilbert spaces at $\bk$ and $\bk+\bG$ are not identical, but related through the unitary mapping $W_{\bG}$. Using the groupoid property we can check that with this definition Wilson loops along the same oriented loop but with different starting points are unitarily related. Furthermore, a gauge transformation or a shift by a reciprocal lattice vector also acts as a unitary transformation, showing that the spectrum is insensitive to these choices, meaning we can consider closed loops $\cC$ as directed curves on the BZ torus in the following. Finally, consider a shift of the real space origin by $\bt$ acting on the coefficients as $e^{i\bk\bt}$. $\cP$ does not change, but the BZ boundary condition changes $W_{\bG} \to W_{\bG} e^{i\bG\bt}$, resulting in $\mathcal{W}_{\mathcal{C}}\to\mathcal{W}_{\mathcal{C}}e^{-i\bG\bt}$, meaning a simple shift in all the phases of the nonzero eigenvalues, $\brac{\phi_{\cC}}\to\brac{\phi_{\cC}-\bG\bt}$ with $\brac{\phi_{\cC}} = -i \log \pars{\sigma\pars{\mathcal{W}_{\mathcal{C}}}\setminus 0}$. This result is easily understood if we regard the curve $\cC$ as the BZ of a 1d system with isolated bands. The phases of the Wilson loop eigenvalues can be identified as the Wannier center positions for occupied bands, and as we showed they indeed transform as real space coordinates in the $\bG$ direction under a shift of the spatial origin.

In the following we establish some symmetry properties of the WL operators, starting with the well known case of time-reversal symmetry. TR symmetry with $\cT = \sigma_y \cK$ implies $\cP_{\bk} = \cT \cP_{-\bk} \cT^{-1} = \sigma_y \cP_{-\bk}^* \sigma_y$, plugging this in (\ref{eqn:closedWL}) for a TR invariant line (such that $-\cC = \cC^{-1}$) we find
\eqn{
\mathcal{W}_{\mathcal{C}} = \pars{W_{2 \bk_0} \cT} \mathcal{W}_{\mathcal{C}}\dag \pars{W_{2\bk_0} \cT}^{-1}
}
where we used the fact that $W_{\bk}$ is a diagonal unitary that commutes with $\sigma_y$. The antiunitary operator $\pars{W_{2\bk_0} \cT}$ squares to $-\id$ same as $\cT$, direct calculation shows that the eigenvectors $\ket{\lambda}$ and $\pars{W_{2\bk_0} \cT}\ket{\lambda}$ are orthogonal, both with eigenvalues $\lambda$, proving the Kramers-degeneracy of WL eigenvalues.

Next we move on to the action of SG symmetries on WL operators, in particular we consider SG operation $\brac{O\middle|\bt}$ represented by $R_{\bk} = e^{-i\pars{O\bk}\bt} U: \mathcal{H}_{\bk}\to\mathcal{H}_{O\bk}$. As we did so far, we keep track of the absolute (not BZ reduced) positions, and transition to a BZ reduced convention when necessary. Using the symmetry property $\cP_{O\bk} = U \cP_{\bk} U\dag$ of the projector we see for any open Wilson line operator
\eqn{
\mathcal{W}_{O \mathcal{C}} =  U \mathcal{W}_{\mathcal{C}} U\dag
}
where $O\cC$ is the image of $\cC$.

As discussed before, the great advantage of our formalism is that the atomic positions are encoded in the boundary conditions obeyed by  operators and symmetry operators take the above simple form. Symmetry operators satisfy the same boundary conditions leading to consistency relations. Consider $R_{\bk+\bG}: \mathcal{H}_{\bk+\bG}\to\mathcal{H}_{O\pars{\bk+\bG}}$, leading to $R_{\bk+\bG} = W_{O\bG} R_{\bk} W_{\bG}^{-1}$ which imposes the constraint on the constant unitary part $e^{-i\pars{O\bG}\bt} U = W_{O\bG} U W_{\bG}^{-1}$. As a result, for closed Wilson loops
\eqn{
\mathcal{W}_{O\mathcal{C}} = e^{i \pars{O \bG} \bt} U \mathcal{W}_{\mathcal{C}} U\dag
}
with the reciprocal lattice vector $\bG$ describing the winding of the loop $\cC$. So the set of phases of the eigenvalues is related as $\brac{\phi_{O\cC}} = \brac{\phi_{\cC} + \pars{O\bG}\bt}$. This property is identical to the symmetry requirement for the projections of the atomic (Wyckoff) positions. The set of atomic positions has to be invariant under the SG symmetry, $\brac{\bx} = \brac{O\bx+\bt}$. Now consider the projection onto $O\bG$, $\brac{\pars{O\bG}\bx} = \brac{\pars{O\bG}\pars{O\bx+\bt}} = \brac{\bG\bx + \pars{O\bG}\bt}$, same requirement as for the phases if we identify phases along a loop with winding $\bG$ with the projections of atomic coordinates along $\bG$. This picture is further supported by considering an atomic insulator with $\cP^{\pars{\bx,l},\pars{\bx',l'}}_{\bk} = \delta^{\bx,\bx'} P_{\bx}^{l,l'}$ where $P_{\bx}^{l,l'}$ is the projector onto the occupied atomic levels at each site, same for all symmetry related sites. In this case $\mathcal{W}_{\mathcal{C}} = \cP W_{-\bG} \cP$ whose eigenvalues are exactly the list of the  projections of the atomic positions of the occupied orbitals onto $\bG$.

Nevertheless, we emphasize that the WL eigenvalues are \emph{not} the same as the projections of the Wannier center position, the latter are obtained integrating the sum of the WL phases for all parallel lines in a given direction, thus contain less information. The Wannier center positions for individual bands are in general ill defined as WL eigenvalues can cross as moving the line $\cC$ in the perpendicular direction. On the other hand, the above symmetry constraints on WL eigenvalues are weaker than those on Wyckoff positions, a symmetry preserving preimage of the projections as a set of 3D points may not exist, such a situation signals a crystalline topological phase distinct from the trivial (atomic insulator) phase.


\section{Alternative derivation of the Chern number formula with $C_n$ symmetry}
\label{sec:Cn}

We briefly review the results of Ref.~\onlinecite{BernevigPGS} constraining the Chern number of insulators with discrete rotational symmetry. We use our formulation for Wilson loop invariants (Appendix~\ref{sec:WL}) and arrive at the same results in a slightly different, but perhaps more transparent way.

As an example we detail the derivation for fourfold ($C_4$) rotations, the other cases can be obtained analogously. We are going to examine the Wilson loop operator along the contractible curve $\cC = \Gamma X M Y$. As this curve is contractible, the enclosed Berry flux modulo $2\pi$ is given by the pseudodeterminant of $\mathcal{W}_{\cC}$, and because of the symmetry properties of $\cF$ this is exactly one quarter of the total flux through the BZ. This can be summarized as $\exp\pars{\frac{2\pi i}{4} C} = \det_+ \mathcal{W}_{\cC}$. We can break up the loop into 4 parts as $\cW_{\cC} = \cW_{\Gamma\leftarrow Y}\cW_{Y\leftarrow M}\cW_{M\leftarrow X}\cW_{X\leftarrow \Gamma}$ and analyze symmetry properties of its parts. We find $\cW_{\Gamma\leftarrow Y} = U \cW_{\Gamma\leftarrow X} U = U \cW_{X\leftarrow \Gamma}^{-1} U\dag$ as they are related by $C_4$ represented by $U$ and reversal of the direction. Here and in the following by $()^{-1}$ we mean the Moore-Penrose pseudoinverse of singular operators, which in this case is the same as the inverse if we restrict both the domain and codomain to the occupied subspace. Similarly $\cW_{Y\leftarrow M} = U\dag W_{-\bG_x} \cW_{M\leftarrow X}^{-1} W_{\bG_x} U$ as now the necessary transformation is a rotation in the opposite direction combined with a translation by a reciprocal lattice vector and reversal of direction.

Substituting this we rearrange the WL operator as
\begin{align}
\cW_{\cC} =& \pars{\cP_{\Gamma} U \cP_{\Gamma}} \cW_{X\leftarrow \Gamma}^{-1} \pars{\cP_X U^{-2} W_{-\bG_x} \cP_X} \cW_{M\leftarrow X}^{-1}\times\nonumber\\
 & \pars{\cP_M W_{\bG_x} U \cP_M} \cW_{M\leftarrow X}\cW_{X\leftarrow \Gamma}.
\end{align}
We inserted band projectors at the end points of the lines without changing the results and introduced parentheses to emphasize the structure. Every term in the product is a unitary when restricted to the occupied subspace at its domain and codomain. Thus the determinant multiplication theorem applies to the pseudodeterminant,
\begin{align}
\det_+ \cW_{\cC} =&\det_+ \pars{\cP_{\Gamma} U \cP_{\Gamma}} \det_+ \pars{\cP_X U^{-2} W_{-\bG_x} \cP_X}\times\nonumber\\ &\det_+ \pars{\cP_M W_{\bG_x} U \cP_M}.
\end{align}
Each of these determinants is exactly the product of the occupied eigenvalues of a $C_4$ or $C_2$ rotation at a given point of the BZ,
\eqn{
\det_+ \cW_{\cC} = \prod_{m\in\textnormal{occ.}} \xi_m^{\Gamma}\pars{C_4} \xi_m^{M}\pars{C_4} \xi_m^{X}\pars{C_4^{-2}}
}
where $\xi_m^{\bk}\pars{O}$ is the rotation eigenvalue of $O$ in band $m$ at momentum $\bk$. Thus we reproduced the result for $C_4$ in Ref.~\onlinecite{BernevigPGS} and Eqn.~\ref{eqn:C4}, results for $C_2$, $C_3$ and $C_6$ symmetry follow analogously:
\begin{align}
e^{2\pi i\frac{C}{2}} &= \prod_{m\in\textnormal{occ.}} \xi_m^{\Gamma}\pars{C_2} \xi_m^{M}\pars{C_2} \xi_m^{X}\pars{C_2^{-1}}\xi_m^{Y}\pars{C_2^{-1}}\\
e^{2\pi i\frac{C}{3}} &= \prod_{m\in\textnormal{occ.}} \xi_m^{\Gamma}\pars{C_3} \xi_m^{K}\pars{C_3} \xi_m^{K'}\pars{C_3^{-2}}\\
e^{2\pi i\frac{C}{6}} &= \prod_{m\in\textnormal{occ.}} \xi_m^{\Gamma}\pars{C_6} \xi_m^{K}\pars{C_6^2} \xi_m^{M}\pars{C_6^{-3}}.
\end{align}

\section{Generalization to topological superconductors}
\label{sec:TSC}
Weak indices are also present in other symmetry classes\cite{Ran, deJuan2014} and we extend our considerations to other cases with $d-1$ dimensional weak indices. In 3D there are analogous 2D $\bbZ$ and $2\bbZ$ indices in classes D and C respectively, these are Chern numbers of the Bogoliubov-de Gennes Hamiltonians and the same reasoning applies as in class A. In other cases, however, the presence of charge conjugation symmetry ($\cC$) has a more important role, in the following we review one example and present the relevant formalism.

In class D we have 1D $\bbZ_2$ indices that serve as $d-1$ dimensional weak indices in a 2D system:
\eqn{
\nu_i = \frac{1}{\pi} \oint \Tr\cA \pmod{2}
}
where the integration contour is along an invariant line on the edge of the BZ $\bG_i/2 + t \epsilon_{ij}\bG_j$, $t\in[0,1]$. Similarly to the weak $\bbZ_2$ in 3D TI's, the value $\nu_i'$ on a parallel invariant line through the $\Gamma$ point is not independent, it is related through the 2d strong index $C\in\bbZ$ such that $C \equiv \pars{\nu_i + \nu_i'} \pmod{2}$. It can be shown that the weak vector $\bG^{\nu} = 1/2 \nu_i \bG_i$ again transforms as $\bk$-space vectors under space-group operations, showing that rhombic and square lattices only allow $(1/2,1/2)$ (in primitive basis) and 3 or 6-fold rotational symmetry does not allow any nontrivial weak vector.

In order to prove the transformation properties of the 1d weak $\bbZ_2$ indices in class D, we have to switch to the BZ periodic convention, as $\Tr_{occ.}\oint\cA$ is only quantized in a periodic basis and gauge. A space group operation $g$ in this basis is represented as $U^{ab} = U_0^{ac}\delta^{cb} e^{-i \bk \delta\bR_b}$ where $\delta\bR_a$ is the lattice vector of the unit cell in which site $a$ of the unit cell with at $\bR = 0$ ends up after the application of $g$. $U_0$ is $\bk$-independent and we set it to the identity without loss of generality. The set $\delta\bR_a$ depends on the choice of the unit cell, and a basis transformation redefining the unit cell has the same form with $\delta\bR_a$ showing the change of unit cell position to which site $a$ is assigned. Now we are in a position to prove two things at once: the 1D $\bbZ_2$ indices in class D transform in a simple fashion under SG operations and are insensitive to the choice of the real space unit cell.

We introduce the band-flattened Hamiltonian, $\mathcal{Q} = \id - 2\cP$, it has the same properties as $H$ except all particle/hole-like bands have energy $\pm 1$. Charge conjugation symmetry $\cC = \tau_x \cK$ imposes $\mathcal{Q}_{\bk} = - \tau_x \mathcal{Q}_{-\bk}^{\ast} \tau_x$, for $\cP$ it means $\cP_{\bk} = \id - \tau_x \cP_{-\bk}^{\ast} \tau_x$. The particle and hole-like states are related by Hermitian conjugation, this in general implies $U_{\bk} = \tau_x U_{-\bk}^{\ast} \tau_x$, for the diagonal form we use this means every $\bR_a$ has to appear twice. This is a consequence of double-counting degrees of freedom, the creation and annihilation operators of the same state must live one the same lattice site. We emphasize that this assumption is not always valid in insulators with effective particle-hole symmetry\cite{deJuan}. We find
\begin{align}
i \oint_0^{\bG} \Tr \pars{\cP U\dag \dee U } = -\frac{i}{2} \oint_0^{\bG} \Tr \pars{U\dag \dee U } =\\ = -\frac{1}{2} \oint_0^{\bG} \sum_a \dee \bk \delta\bR_a =
 -\frac{1}{2} \sum_a \bG \delta\bR_a
\end{align}
where $\bG$ is the reciprocal lattice vector along which the integration contour for $\Tr_{occ.}\oint\cA$ is oriented. Because of the doubling of orbitals the right hand side of the equation is always an integer multiple of $2\pi$. Comparing with (\ref{eqn:intA}) we see that a change of the unit cell or a space group operation (with the appropriate transformation on $\bk$-space) does not change the value of $\Tr_{occ.}\oint\cA$ which is only defined modulo $2\pi$.

\section{No Constraints on Strong TI's}
\label{sec:strongTI}
In general, a topological phase from Kitaev's periodic table, protected by a global symmetry of the 10 AZ classes is robust against breaking lattice symmetry, such as strong TI in 3D. If a phase is compatible with a group $G$, then it is also compatible with any space group that is a subgroup of $G$. This is simply true because the topological protection doesn't rely on $G$, all the symmetry restrictions in $G$ can do is to rule out certain phases in the original classification. A subgroup can't rule out more phases, as it poses less restrictions. Of course it is possible to have phases that are protected by $G$ (and the global symmetry), then breaking $G$ down to a subgroup can either allow more phases or protect less. For example, as we saw, nonsymmorphic symmetry can give interesting results about weak indices, as they rely on the translation part of the space group for protection.

As every crystallographic space group is a subgroup of either SG \#229 (Im3m) or \#191 (P6/mmm), finding examples of strong TI's in both of these crystal structures proves that crystal symmetry cannot forbid strong TI's: starting from either of these maximally symmetric examples weakly breaking some of the lattice symmetries one can produce a system with any SG without leaving the strong TI phase.

In our tight-binding examples we have a single site per unit cell with two orbitals, four bands in total. One of the orbitals is a spinful s-orbital, transforming under rotations with the canonical SU(2) representation and even under inversion. The other orbital transforms the same way under proper rotations, but odd under inversion, such orbitals naturally arise through crystal-field splitting of p-orbitals in a spin-orbit coupled ion. We introduce the Pauli matrices $\tau$ to act on the space of the two orbitals, now proper rotations by angle $\mathbf{n}$ are represented as $\exp\pars{\frac{i}{2} \mathbf{n}\cdot \boldsymbol{\sigma}}$, inversion as $\tau_z$ and time-reversal as $\cT = \sigma_y\cK$. Both minimal models have the same form that guarantees that they are invariant under the full symmetry group:
\eqn{
H(\bk) = \sum_{\boldsymbol{\delta}}\brac{\sin\pars{\bk\cdot\boldsymbol{\delta}} \pars{\boldsymbol{\delta}\cdot\boldsymbol{\sigma}}\tau_x + \pars{m-\cos\pars{\bk\cdot\boldsymbol{\delta}}}\tau_z}
}
where the sum runs over nearest neighbor vectors. By tuning $m$ we can enter the strong TI phase, this can be easily checked by counting inversion eigenvalues.

This result is expected based on the band inversion picture, the $\Gamma$ point is always of maximal symmetry, it is possible to have a band inversion only at the $\Gamma$ point, resulting in a strong TI with trivial weak indices. We can also rationalize this result from the effective field theory point of view. The strong TI phase is characterized by the topological $\theta$ term in the long wavelength electromagnetic action, a theory that possesses continuous translation an rotation symmetries. While a microscopic theory with full Galilean invariance is not possible, we showed that the maximally symmetric crystal structures are all compatible with this emergent behavior.

\end{document}